# Finite Spectral Quantum Field Theory

*Dedicated to the memories of my PhD thesis advisor and friend, the late Christian Fronsdal, who was passionate about sailing his self-built trimaran and playing Go as well as doing mathematical physics*

A. D. Alhaidari

*Saudi Center for Theoretical Physics, P.O. Box 32741, Jeddah 21438, Saudi Arabia*

Email: haidari@sctp.org.sa

**Abstract:** Using the spectral properties of orthogonal polynomials, we introduce an algebraic version of quantum field theory for elementary particles. Closed-loop integrals in the Feynman diagrams for computing transition amplitudes are divergence-free. Consequently, the theory is finite and no renormalization scheme is required.

## 1. Introduction

Quantum field theory (QFT) was developed almost a century ago to describe elementary particles, like electrons and neutrons, their interaction with each other and with their environment. Early foundational work was carried out in the mid-1920s by prominent scientists like Heisenberg, Born, and Jordan who contributed to the quantization of the electromagnetic field. On the other hand, 1927 is the year most often cited for the formal introduction of the theory by Dirac where he coined the name Quantum Electrodynamics (QED), which is considered the first successful QFT. However, the theory had to be augmented by ad hoc rules (called renormalization and regularization) for computing physical processes because all such computations are plagued with infinities. Out of the many renormalization schemes aimed at eliminating these infinities that arise in perturbative calculations, two are notable. The Epstein-Glaser renormalization, developed by Epstein and Glaser in the 1970s [1], is a scheme based on causal perturbation theory. It works directly with the S-matrix and its coefficients, which are treated as operator-valued distributions in position space. The BPHZ renormalization, a refinement of the Bogoliubov-Parasiuk method [2] by Hepp [3] and Zimmermann [4], is a widely used momentum-space renormalization scheme by subtracting divergences within the perturbative expansion. However, up to date, all efforts to develop a consistent QFT free of divergencies were not fully satisfactory. In this work, we attempt at a resolution of this shortcoming by introducing an algebraic version of QFT based on the spectral properties of orthogonal polynomials that is inherently finite and thus does not require renormalization altogether. That is, a theory in which closed-loop integrals in the Feynman diagrams for calculating transition amplitudes are divergence-free.

We assume basic knowledge of QFT and thus will not dwell on a review of the theory. Interested readers may consult any of the classic literature on the subject such as those in Refs. [5-7]. Additionally, we will not assume advanced knowledge in the theory of orthogonal polynomials. It is sufficient that the reader be familiar with the properties of well-known orthogonal polynomials such as the Hermite, Laguerre, Chebyshev, Gegenbauer, ...etc. At least, one must know where to find such properties by looking at any of the numerous books on special functions and orthogonal polynomials like those listed in [8-11]. For the purpose of the present development, one should be aware that an orthogonal set of polynomials has an associated weight function, satisfies a three-term recursion relation and an orthogonality.

The discrete expansion of the plane wave $e^{i\vec{k}\cdot\vec{r}}$ in the linear momentum representation of quantum fields in conventional QFT (see Appendix A) suggests a discrete spectral decomposition of the free quantum field in the energy space. The new representation has an algebraic structure built on orthogonal energy polynomials. For example, the differential free wave equation in conventional QFT becomes an algebraic three-term recursion relation satisfied by the orthogonal polynomials associated with the given quantum field. Therefore, instead of solving a differential wave equation, one can simply solve an algebraic equation. Next, we show that the equal-time commutation relation of the quantum field and its canonical conjugate in this theory is a direct consequence of the orthogonality of the associated polynomials and the completeness of the "spectral modes" of the quantum field. Additionally, we show how the basic singular two-point function, whose time ordering leads to the Feynman propagator, is compatible with micro-causality.

Nonetheless, a free QFT (i.e., a theory without interaction) does not carry much physical significance beyond the obvious requirement of being physically and mathematically consistent and preferably elegant. Hence, we come to our most important findings where we show that closed loop integrals in the Feynman diagrams used for calculating transition amplitudes in this theory are finite. We verify this claim in a typical Yukawa-type interaction model for few sample diagrams. This property relies predominantly on the orthogonality property of the associated polynomials. We give a numerical illustration showing that the finiteness property of this model continues to higher order loops. It remains to be seen whether this remarkable property is maintained in other physically relevant models.

To make the presentation clear and simple, we limit our detailed description to a scalar particle in 3+1 Minkowski space-time and adopt the conventional relativistic units $\hbar = c = 1$. Moreover, we limit our detailed treatment to the positive-energy component of the quantum field. Moreover, in several Appendices we give a brief treatment of the Dirac spinor, massless and massive vector fields in this theory. However, in an introductory article like this one, it is impossible to address all relevant issues of the theory that were investigated by countless researchers for over a century of conventional QFT. For example, we leave the issue of relativistic invariance of the theory to a future investigation. Nonetheless, under the assumption that the theory is invariant, we show that causality is respected.

## 2. Scalars in spectral QFT

Motivated by the findings in Appendix A where the quantum field in conventional QFT could be written as an infinite sum of discrete excitations, we introduce an algebraic version of QFT in the energy space. In this theory, the positive energy component of the scalar quantum field in 3+1 Minkowski space-time is represented by the following Fourier expansion in the energy

$$\Psi(t,\vec{r}) = \int dE \sqrt{2E}\, e^{-iEt} \sum_{n=0}^{\infty} a_n(E)\, \phi_n(\vec{r}), \qquad (1)$$

where $E^2 \geq M^2$ and on the physical subspace $a_n(E)|\text{phys}\rangle = p_n(z)\left[a_0(E)|\text{phys}\rangle\right]$ making $p_0 = 1$ with $z$ being a real dimensionless energy-dependent parameter called the *spectral*

*parameter*. $\{p_n(z)\}_{n=0}^{\infty}$ is a complete set of *real* functions of *z* whereas $a_0(E)$ is an operator-valued object (the annihilation operator) that satisfies the following commutation algebra

$$\left[a_0(E),a_0^{\dagger}(E')\right]=a_0(E)a_0^{\dagger}(E')-a_0^{\dagger}(E')a_0(E)=f(E)\delta(E-E'), \tag{2}$$

and $f(E)$ is a positive definite function of the energy to be determined below by canonical quantization. $\{\phi_n(\vec{r})\}_{n=0}^{\infty}$ is a complete set of spatial functions that form an invariant subspace in the domain of the 3D Laplacian operator. That is, $\vec{\nabla}^2\phi_n \in \{\phi_m\}_{m=0}^{\infty}$. Specifically, they are required to satisfy the following differential relation

$$-\vec{\nabla}^2\phi_n(\vec{r})=\lambda^2\left[\alpha_n\phi_n(\vec{r})+\beta_{n-1}\phi_{n-1}(\vec{r})+\beta_n\phi_{n+1}(\vec{r})\right], \tag{3}$$

where $\lambda$ is a real universal constant of inverse length dimension (a universal scale/mass)*. The coefficients $\{\alpha_n,\beta_n\}_{n=0}^{\infty}$ are real dimensionless parameters such that $\beta_n\neq 0$ for all *n* and by definition $\beta_{-1}:=0$. Using (3) in the free Klein-Gordon wave equation, $\left(\partial_t^2-\vec{\nabla}^2+M^2\right)\Psi(t,\vec{r})$ $=0$, results in the following algebraic relation

$$z\,p_n(z)=\alpha_n p_n(z)+\beta_{n-1}p_{n-1}(z)+\beta_n p_{n+1}(z), \tag{4}$$

for $n=1,2,3,...$ and with $z=(E^2-M^2)/\lambda^2=\vec{k}^2/\lambda^2$. This equation is a symmetric three-term recursion relation that makes $\{p_n(z)\}$ a sequence of polynomials in *z* (called the *spectral polynomials*) with the two initial values $p_0(z)=1$ and $p_1(z)$ a two-parameter linear function of *z*. Now, Eq. (4) has two linearly independent polynomial solutions. We choose the solution with the initial values $p_0(z)=1$ and $p_1(z)=(z-\alpha_0)/\beta_0$.† The spectral theorem (a.k.a. Favard theorem) [8-11] states that the polynomial solutions of the recursion (4), with the recursion coefficients $\{\beta_n^2\}$ being positive definite, must satisfy the following orthogonality relation

$$\int\rho(z)p_n(z)p_m(z)dz=\delta_{n,m}, \tag{5}$$

with $\rho(z)$ being the associated weight function, which is positive definite and will be related to $f(E)$ of Eq. (2) by canonical quantization. All elements of the theory including $p_n(z)$, $\phi_n(\vec{r})$, $\alpha_n$, $\beta_n$, $a_0(E)$, $\rho(z)$ transform in special manner under the action of the Poincaré transformation (space-time translation and rotation) such that relativistic invariance and covariance of the theory is maintained. This will be detailed in a future investigation.

---

* The physical interpretation of this universal parameter is to be determined. It could be that $\lambda^4$ is the energy density of the vacuum of our theory, or that $\lambda^2$ is proportional to the cosmological constant of space-time, or that $\lambda$ is the mass of the Higgs particle, etc.

† Changing the initial values to $p_0(z)=1$ and $p_1(z)=c_1z-c_2$ where $c_1$ and $c_2$ are real parameters such that $c_1\neq 1/b_0$ and/or $c_2\neq a_0/b_0$ (also, $c_1\neq 0$) will change the polynomials and their weight function $\rho(z)$ in the orthogonality relation (5). In fact, such a change could also create an internal structure for the quantum field in which $\rho(z)$ gets augmented by a number of delta functions $\delta(z-\hat{z})$ at discrete negative values of *z* within the interval $-M^2/\lambda^2\leq\hat{z}<0$ and thus the orthogonality will have a finite discrete sum in addition to the integral (5) (see Ref. [13] for details).

The fundamental algebraic relation (4), which is equivalent to the free Klein-Gordon wave equation, is the reason behind the spectral and algebraic structure of the theory. In fact, postulating the three-term recursion relation (4) eliminates the need for specifying a free field wave equation. Furthermore, once the set of spectral polynomials $\{p_n(z)\}$ is given then all physical properties of the corresponding particle are determined. This is so because once $\{p_n(z)\}$ is given then $\rho(z)$ and $\{\alpha_n,\beta_n\}$ are identified leading to $a_0(E)$ and $\{\phi_n(\vec{r})\}$ as shown by Eq. (2) and Eq. (3), respectively. Therefore, the spectral polynomials carry a faithful representation of the quantum field and constitute the only essential ingredient in the theory. All physical processes, including transition amplitudes, are fully determined once changes to these polynomials are identified. It is worthwhile noting that in the linear momentum ($k$-space) representation of conventional QFT, the scalar quantum field could be written in a form similar to (1). We verify this assertion in Appendix A.

Now, the conjugate quantum field $\overline{\Psi}(t,\vec{r})$ is obtained from (1) by complex conjugation and the replacement $\phi_n(\vec{r})\mapsto\overline{\phi}_n(\vec{r})$ where the pair $\{\phi_n(\vec{r}),\overline{\phi}_n(\vec{r})\}$ satisfies the following orthogonality and completeness properties

$$\left\langle\phi_n(\vec{r})\middle|\overline{\phi}_{n'}(\vec{r})\right\rangle=\left\langle\overline{\phi}_n(\vec{r})\middle|\phi_{n'}(\vec{r})\right\rangle=\delta_{n,n'}\,, \tag{6a}$$

$$\sum_{n=0}^{\infty}\phi_n(\vec{r})\overline{\phi}_n(\vec{r}')=\sum_{n=0}^{\infty}\overline{\phi}_n(\vec{r})\phi_n(\vec{r}')=\delta^3(\vec{r}-\vec{r}')\,. \tag{6b}$$

Therefore, we write $\overline{\Psi}(t,\vec{r})$ as

$$\overline{\Psi}(t,\vec{r})=\int dE\sqrt{2E}\,e^{\mathrm{i}Et}\sum_{n=0}^{\infty}a_n^{\dagger}(E)\overline{\phi}_n(\vec{r})\,. \tag{7}$$

Using the commutators of the creation/annihilation operators shown above as Eq. (2) and noting that $2EdE=\lambda^2dz$, we can write

$$\left[\Psi(t,\vec{r}),\overline{\Psi}(t',\vec{r}')\right]=\sum_{n,m=0}^{\infty}\phi_n(\vec{r})\overline{\phi}_m(\vec{r}')\left[\int e^{-\mathrm{i}E(t-t')}\lambda^2 f(E)p_n(z)p_m(z)\,dz\right]. \tag{8}$$

The orthogonality (5) and the completeness (6b) turn this equation with $t=t'$ into

$$\left[\Psi(t,\vec{r}),\overline{\Psi}(t,\vec{r}')\right]=\delta^3(\vec{r}-\vec{r}')\,, \tag{9}$$

provided that we take $\lambda^2 f(E)=\rho(z)$, which makes the commutator (2) read as follows

$$\left[a_0(E),a_0^{\dagger}(E')\right]=\lambda^{-2}\rho(z)\delta(E-E')\,. \tag{10}$$

Moreover, it is straightforward to write

$$\left[\Psi(t,\vec{r}),\Psi(t,\vec{r}')\right]=\left[\overline{\Psi}(t,\vec{r}),\overline{\Psi}(t,\vec{r}')\right]=0\,. \tag{11}$$

In the canonical quantization of fields [5-7], equations (9) and (11) give the canonical conjugate to $\Psi(t,\vec{r})$ as $\Pi(t,\vec{r})=\mathrm{i}\overline{\Psi}(t,\vec{r})$. Furthermore, in analogy with conventional QFT [5-7], we can write Eq. (8) as

$$\left[\Psi(t,\vec{r}),\bar{\Psi}(t',\vec{r}')\right]=\mathrm{i}\,\Delta(t-t',\vec{r}-\vec{r}'),\tag{12}$$

giving the singular two-point function $\Delta(t-t',\vec{r}-\vec{r}')$ as follows

$$\mathrm{i}\,\Delta(t-t',\vec{r}-\vec{r}')=\sum_{n,m=0}^{\infty}\phi_n(\vec{r})\bar{\phi}_m(\vec{r}')\left[\int e^{-\mathrm{i}E(t-t')}\rho(z)p_n(z)p_m(z)\,dz\right].\tag{13}$$

Additionally, Eq. (9) and Eq. (13) give: $\mathrm{i}\,\Delta(0,\vec{r}-\vec{r}')=\delta^3(\vec{r}-\vec{r}')$. Micro-causality, which implies that nothing travels faster than the speed of light, requires that $\left[\Psi(t,\vec{r}),\bar{\Psi}(t',\vec{r}')\right]=0$ for space-like separation where $(t-t')^2-\left|\vec{r}-\vec{r}'\right|^2<0$. Assuming relativistic invariance of $\Delta(t-t',\vec{r}-\vec{r}')$ and choosing a reference frame where $t=t'$, a space-like line element must have $\left|\vec{r}-\vec{r}'\right|^2>0$ and thus $\Delta(0,\vec{r}-\vec{r}')=-\mathrm{i}\,\delta^3(\vec{r}-\vec{r}')=0$. In other words, the theory preserves causality.

The vacuum expectation of the time ordered $\Delta(t-t',\vec{r}-\vec{r}')$ gives the Feynman propagator $\Delta_F(t-t',\vec{r}-\vec{r}')$. Therefore, the spectral polynomials $\{p_n(z)\}$ are the only elements needed to define the free sector of this QFT whereas the set of spatial functions $\{\phi_n(\vec{r})\}$ are determined by solving Eq. (3) using only knowledge of the recursion coefficients $\{\alpha_n,\beta_n\}$ associated with $\{p_n(z)\}$ and the boundary conditions compatible with the two initial values $p_0(z)$ and $p_1(z)$. However, explicit construction of $\{\phi_n(\vec{r})\}$ is not necessary for determining the dynamics (e.g., computing transition amplitudes) making the spectral polynomials the only required and sufficient elements of the theory. This will be clearly demonstrated in the following section.

A real (neutral) scalar particle in 3+1 space-time is represented by the quantum field $\frac{1}{\sqrt{2}}\left[\Psi(t,\vec{r})+\lambda^{-1}\bar{\Psi}(t,\vec{r})\right]$ with $\bar{\phi}_n(\vec{r})\propto\lambda\phi_n^*(\vec{r})$. On the other hand, a complex (charged) scalar particle is represented by a quantum field whose positive-energy component is $\frac{1}{\sqrt{2}}\left[\Psi_+(t,\vec{r})+\Psi_-^\dagger(t,\vec{r})\right]$ and its negative-energy component is $\frac{1}{\sqrt{2}}\left[\bar{\Psi}_+(t,\vec{r})+\bar{\Psi}_-^\dagger(t,\vec{r})\right]$ where $\Psi_\pm(t,\vec{r})$ are identical to (1) but with associated spectral polynomials $\{p_n^\pm(z)\}$ along with their recursion coefficients $\{\alpha_n^\pm,\beta_n^\pm\}$ and weight functions $\rho^\pm(z)$, spatial functions $\{\phi_n^\pm(\vec{r}),\bar{\phi}_n^\pm(\vec{r})\}$, and the annihilation operators $a_0^\pm(E)$ that satisfy $\left[a_0^s(E),a_0^{s'\dagger}(E')\right]=\delta_{s,s'}\lambda^{-2}\rho^s(z)\delta(E-E')$ where $s$ and $s'$ stand for $\pm$.

The particle propagator is an operator that takes the particle from the space-time point $(t',\vec{r}')$ to $(t,\vec{r})$ with $t>t'$. In other words, the particle is created from vacuum at $(t',\vec{r}')$ then annihilated later at $(t,\vec{r})$. In the present algebraic theory, the equivalent "*spectral propagator*" is determined by the two-point function $\langle 0|a_n(E)a_{n'}^\dagger(E')|0\rangle$ where $|0\rangle$ is the vacuum. Now, since $a_n(E)|0\rangle=0$, then we can write

$$\langle 0|a_n(E)a_{n'}^\dagger(E')|0\rangle=\langle 0|\left[a_n(E),a_{n'}^\dagger(E')\right]|0\rangle=\lambda^{-2}\rho(z)p_n(z)p_{n'}(z')\delta(E-E').\tag{14}$$

We designate the spectral propagator by $K_{n,n'}(z,z')$. It takes $p_{n'}(z') \mapsto p_n(z)$ [more precisely, $\sqrt{\rho(z')}p_{n'}(z') \mapsto \sqrt{\rho(z)}p_n(z)$]. That is, the spectral propagator must satisfy

$$\sum_{n'} \int dz' K_{n,n'}(z,z') \left[ \sqrt{\rho(z')}p_{n'}(z') \right] = \sqrt{\rho(z)}p_n(z). \tag{15}$$

The orthogonality (5) shows that the representation $K_{n,n'}(z,z') = \sqrt{\rho(z')\rho(z)}\, p_n(z')p_{n'}(z)$ satisfies (15). This is also suggested by the two-point function of equation (14) that could be rewritten as follows

$$\langle 0 | a_n(E) a^\dagger_{n'}(E') | 0 \rangle = \lambda^{-2} K_{n,n'}(z,z') \delta(E-E'), \tag{16}$$

Consequently, $K_{n,n'}(z,z')$ is the propagator for individual *spectral modes*[‡] of the quantum field ($\text{mode}\, n' \mapsto \text{mode}\, n$) leaving all other modes unaffected. This is unlike the conventional theory where the entire quantum field with all of its spectral modes are propagated. Using the completeness of the spectral polynomials that reads $\sum_n p_n(z)p_n(z') = g(z)\delta(z-z')$ where $g(z) \propto 1/\rho(z)$, one can show that $\sum_n K_{n,n}(z,z') \propto \delta(z-z')$. Moreover, using the orthogonality (5), it is evident that $\int K_{n,n'}(z,z)dz = \delta_{n,n'}$. Additionally, it has the exchange symmetry $K_{n',n}(z,z') = K_{n,n'}(z',z)$. Energy conservation, on the other hand, which is evident by the presence of the Dirac delta function $\delta(E-E')$ in the propagator equation (14), implies that $K_{n,n'}(z,z') \mapsto K_{n,n'}(z,z) := K_{n,n'}(z) = \rho(z)\, p_n(z) p_{n'}(z)$. Now since spectral modes cannot propagate at rest mass energy where $z=0$, then we must require that $K_{n,n'}(0)=0$. Accordingly, spectral polynomials are physically restricted to those with weight functions that satisfy $\rho(0)=0$. This includes:

- The Laguerre polynomial, $L_n^\nu(z)$, with $\nu > 0$ where $\rho(z) \propto z^\nu e^{-z}$,
- The odd Hermite polynomial, $\frac{1}{\sqrt{z}} H_{2n+1}(\sqrt{z})$, where $\rho(z) \propto \sqrt{z}\, e^{-z}$,
- The continuous dual Hahn polynomial [10], $S_n(z;a,b,c)$, with $\{a,b,c\}$ positive and where $\rho(z) \propto x^{-1} \left| \Gamma(a+ix)\Gamma(b+ix)\Gamma(c+ix)/\Gamma(2ix) \right|^2$ and $z = x^2$,
- The odd Meixner-Pollaczek polynomial [8], $\frac{1}{\sqrt{z}} P^{(\nu)}_{2n+1}(\sqrt{z};\theta)$, with $\nu > 0$ and $\theta = \pi/2$ where $\rho(z) \propto \sqrt{z} \left| \Gamma(\nu + i\sqrt{z}) \right|^2$,[§]
- …etc.

---

[‡] The $n^{\text{th}}$ spectral mode is a space-time wave function $e^{-iEt}\phi_n(\vec{r})$ with energy normalization factor $\sqrt{\rho(z)}\, p_n(z)$. Typically, it is oscillatory over the whole space with *n* nodes and asymptotically decaying amplitude.

[§] One can show that $P^{(\nu)}_{2n}(x;\pi/2) = (-1)^n \frac{2^{2n}}{(2n)!} S_n(x^2;\nu,0,\frac{1}{2})$ and $P^{(\nu)}_{2n+1}(x;\pi/2) = (-1)^n \frac{2^{2n+1}}{(2n+1)!} x\, S_n(x^2;\nu,1,\frac{1}{2})$, which is analogous to the Hermite/Laguerre polynomial relation shown in Appendix A by Eq. (A3).

In the following section, we show how this propagator enters in the calculation of the transition amplitudes via the Feynman diagrams.

## 3. Interaction in spectral QFT

In this section, we show how to calculate the transition amplitude $\langle \text{out} | \text{in} \rangle$ in a generic Yukawa type interaction model. We consider the interaction Lagrangian $\mathscr{L}_I = \eta \bullet \left[ \Phi(\bar{\mathcal{X}}\mathcal{X}) \right]$ and choose $\Phi$ to be a scalar, $\mathcal{X}$ a spinor (see Appendix B for the construction of the spinor quantum field in this theory), and $\eta$ a dimensionless coupling tensor of rank three that couples individual spectral field modes. This interaction Lagrangian resembles that of QED where the scalar field is replaced by the massless vector field (the photon, see Appendix C). If we designate $\{q_n(z)\}$ as the set of spectral polynomials associated with the spinor component $\mathcal{X}$, then we can write the corresponding spinor annihilation operators $b_n(E) = b_0(E) q_n(z)$ and their anti-commutation relation $\left\{ b_0(E), b_0^\dagger(E') \right\} = \lambda^{-2} \omega(z) \delta(E - E')$ where $z = (E^2 - M_{\mathcal{X}}^2)/\lambda^2$ and $\omega(z)$ is the weight function associated with $\{q_n(z)\}$. For simplicity, we consider neutral particles and a single spinor component where we can write

$$\begin{aligned} \mathscr{L}_I(t,\vec{r}) = \sum_{n,m,k=0}^{\infty} \eta_n^{m,k} \Big[ \iiint dE dE' dE'' \sqrt{2E\,2E'\,2E''}\, e^{-\mathrm{i}(E - E' + E'')t} \\ a_n(E) b_m^\dagger(E') b_k(E'') \phi_n(\vec{r}) \bar{\vartheta}_m(\vec{r}) \vartheta_k(\vec{r}) \Big] + \text{Hermitian conjugate} \end{aligned} \tag{17}$$

where $\vartheta_n = \begin{pmatrix} \vartheta_n^+ \\ \vartheta_n^- \end{pmatrix}$ is the four-component spinor basis functions that satisfy Eq. (B4) and Eq. (B9) in Appendix B. The coupling tensor $\eta$ is a function of all recursion coefficients associated with the spectral polynomials of the interacting fields. Moreover, as the indices of $\eta$ tend to infinity, its asymptotics go like $\eta \approx j^{-\tau_j}$ for each index $j$, where $\tau_j > 1$. The fermionic structure of the present interaction model dictates that it is antisymmetric in the fermionic indices. That is, $\eta_n^{m,k} = -\eta_n^{k,m}$.

As an illustrative example of calculating the transition amplitudes, we consider the first order (one-loop) correction to the free propagator (i.e., self-energy). Figure 1 represents the associated Feynman diagram in this model where the propagator for the scalar (spinor) is represented by a dashed (solid) curve, respectively. From this point forward, we designate the spectral propagator for the scalar as $\square_{n,m}(z) := K_{n,m}(z) = \rho(z) p_n(z) p_m(z)$ with $\lambda^2 z = E^2 - M_\Phi^2$ and that for the spinor as $\Delta_{n,m}(z) := \omega(z) q_n(z) q_m(z)$ with $\lambda^2 z = E^2 - M_{\mathcal{X}}^2$. Since the spectral polynomials carry a faithful representation of the quantum field, then we can adopt the notation $\langle \text{out} | \text{in} \rangle \mapsto \langle p_m(z) | p_n(z) \rangle$ for the amplitude shown in the Figure with $\lambda^2 z = E^2 - M_\Phi^2$. Moreover, the spinor propagator for the top part of the closed loop is $\Delta_{i,k}(z') = \omega(u) q_i(u) q_k(u)$ with $u = z' = (E'^2 - M_{\mathcal{X}}^2)/\lambda^2$. Conservation of energy in the figure dictates that $E'' = E + E'$ making the spinor propagator for the bottom part of the loop read $\Delta_{l,j}(z'') = \omega(w) q_l(w) q_j(w)$ with

$$w = z'' = [(E+E')^2 - M_{\mathcal{X}}^2]/\lambda^2 = u + z + (M_\Phi/\lambda)^2 + 2\sqrt{[u+(M_{\mathcal{X}}/\lambda)^2][z+(M_\Phi/\lambda)^2]}\,, \quad (18)$$

In the algebraic system introduced in Ref. [12], the spectral parameter addition (18) is written as $w = u \oplus z$ with each parameter having and assigned parity. Since spectral propagation is possible only for real momenta $\vec{k}$ (i.e., $E^2 > M^2$ and thus $z > 0$) then in Figure 1, all propagators for real and virtual particles must have positive spectral parameters $z$, $z'$, and $z''$ with properly assigned parities. Moreover, under these conditions, the expression under the square root in Eq. (18) is positive definite.

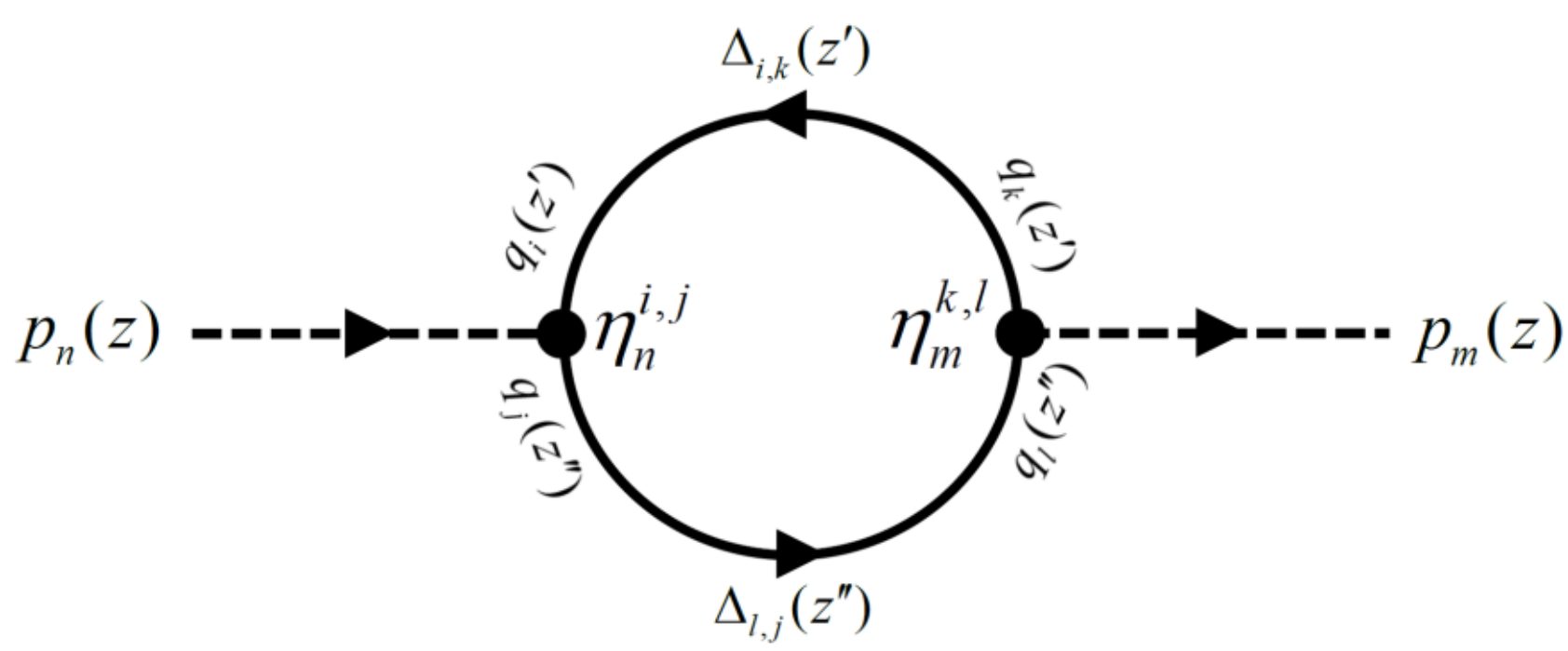

**Fig. 1**: Feynman diagram for the first order (single loop) correction to the scalar propagator (i.e., self-energy) in the model $\mathscr{L}_I = \eta \bullet \left[\Phi(\bar{\mathcal{X}}\mathcal{X})\right]$

To compute the amplitude $\langle p_m(z)|p_n(z)\rangle$ of Figure 1, we follow the Feynman diagram rules. We integrate over all possible values of $u$ and sum over all possible indices (polynomial degrees). That is, the spectral propagator for the scalar particle to first order becomes

$$\begin{aligned}\langle p_m(z)|p_n(z)\rangle &\approx \square_{n,m}(z) + \sum_{i,j,k,l,n',m'=0}^{\infty} \square_{n,n'}(z)\left[\eta_{n'}^{i,j}\eta_{m'}^{k,l}\int \Delta_{i,k}(u)\Delta_{l,j}(u\oplus z)\,du\right]\square_{m',m}(z)\\ &= \square_{n,m}(z)\left\{1 + \sum_{i,j,k,l,n',m'=0}^{\infty} \eta_{n'}^{i,j}\eta_{m'}^{k,l}\,\square_{n',m'}(z)\int \Delta_{i,k}(u)\Delta_{l,j}(u\oplus z)\,du\right\}\end{aligned} \quad (19)$$

The integral in this amplitude is one of the "fundamental SAQFT integrals" introduced in Ref. [13] and denoted by $\zeta_{i,j}^{q,q}(z)$ for $i = k$ and $j = l$ (i.e., for "monochrome propagation"). In general, transition amplitudes are written as $\langle \text{out}|\text{in}\rangle = \langle \text{out}|\text{in}\rangle_0^i + \langle \text{out}|\text{in}\rangle_1^j + \langle \text{out}|\text{in}\rangle_2^k + ...$ where $\langle \text{out}|\text{in}\rangle_n^m$ is the term (or sum of terms) in the perturbation series that corresponds to an amputated and topologically distinct Feynman diagram with $n$ vertices and $m$ loops.[**] Therefore, in (19) $\langle p_m(z)|p_n(z)\rangle_0^0 = \square_{n,m}(z)$ and $\langle p_m(z)|p_n(z)\rangle_{n\geq 1}^0 = 0$ whereas $\langle p_m(z)|p_n(z)\rangle_2^1$ is the second term in (19) without the sum over the factor $\square_{n',m'}(z)$. That is,

[**] $m$ loops mean an integration over $m$ independent spectral parameters.

$$\begin{aligned}\langle p_m(z)|p_n(z)\rangle_2^1 &= \sum_{i,j,k,l=0}^{\infty} \eta_n^{i,j}\eta_m^{k,l} \int \Delta_{i,k}(u)\Delta_{l,j}(u\oplus z)\,du \\ &= \sum_{i,j,k,l=0}^{\infty} \eta_n^{i,j}\eta_m^{k,l} \int \omega(u)\omega(u\oplus z) q_i(u) q_k(u) q_l(u\oplus z) q_j(u\oplus z)\,du\end{aligned} \tag{20}$$

The finiteness of transition amplitudes, like (20), is two-fold; one for the integral and another for the sum. The first, is the finiteness of the integral, which is guaranteed by the orthogonality (5) of the spectral polynomials††. In fact, we have shown in [13] (see also Appendix E below) that the value of $\zeta_{i,j}^{q,q}(z)$ is within the interval [0,1] for all spectral polynomials satisfying the orthogonality (5). Moreover, its value goes to zero fast enough if any of the indices *i* or *j* go to infinity. The second issue in the finiteness of the transition amplitude is the convergence of the series, which depends on the asymptotics of the coupling tensor $\eta$ and the sign signature of its components.

In the integral of Eq. (20), the spectral parameter *u* assumes values that correspond to all possible ranges of the energy $E'$ (i.e., $E' \geq M_{\mathcal{X}}$ and $E' \leq -M_{\mathcal{X}}$). Therefore, the integral runs over all positive values of *u* from 0 to $+\infty$ since $u = (E'^2 - M_{\mathcal{X}}^2)/\lambda^2$ . In the following section, we give a numerical illustration of the finiteness of the transition amplitude (20) and show that monochrome propagation [i.e., $K_{n,n'}(z) \mapsto K_{n,n}(z)$] enhances the value of the transition amplitude and its convergence. Additionally, we compute a truncated version of the series in (19) demonstrating converges as the number of terms increases.

The second order (two-loop) correction to the scalar propagator contains several diagrams one of which is shown below as Figure 2. To compute this second order correction, we integrate over all possible values of *u* and *w* and sum over all possible indices $\{i,j,k,l\}$ and $\{r,s,t,a,b,c\}$ . To simplify, we assume monochrome propagation where the degrees of the spectral polynomials are preserved in propagation. That is, $\square_{n,m}(z) \mapsto \square_{n,n}(z)$ and $\Delta_{n,m}(z) \mapsto \Delta_{n,n}(z)$. In the diagram of Figure 2, this is equivalent to the replacement $\eta_r^{s,t} \mapsto \eta_r^{i,l}$ and $\eta_a^{b,c} \mapsto \eta_r^{k,j}$ . Accordingly, we obtain the following second order contribution to the self-energy of the scalar propagator

$$\begin{aligned}\langle p_m(z)|p_n(z)\rangle_4^2 &= \sum_{i,j,k,l,r=0}^{\infty} \eta_n^{i,j}\eta_m^{k,l}\eta_r^{i,l}\eta_r^{k,j} \times \\ &\iint \square_{r,r}(w)\Delta_{i,i}(u)\Delta_{j,j}(u\oplus z)\Delta_{l,l}(u\oplus w)\Delta_{k,k}(u\oplus w\oplus z)\,du\,dw \\ &= \sum_{i,j,k,l,r=0}^{\infty} \eta_n^{i,j}\eta_m^{k,l}\eta_r^{i,l}\eta_r^{k,j} \iint \rho(w)\omega(u)\omega(u\oplus z)\omega(u\oplus w)\omega(u\oplus w\oplus z) \times \\ &\quad p_r^2(w) q_i^2(u) q_j^2(u\oplus z) q_l^2(u\oplus w) q_k^2(u\oplus w\oplus z)\,du\,dw\end{aligned} \tag{21}$$

Where

$$\begin{aligned}\lambda^2(u\oplus w\oplus z) &= (E_u + E_w + E_z)^2 - M_{\mathcal{X}}^2 = \lambda^2(u+w+z) + 2M_\Phi^2 \\ &+2\sqrt{(\lambda^2 u + M_{\mathcal{X}}^2)(\lambda^2 w + M_\Phi^2)} + 2\sqrt{(\lambda^2 w + M_\Phi^2)(\lambda^2 z + M_\Phi^2)} + 2\sqrt{(\lambda^2 z + M_\Phi^2)(\lambda^2 u + M_{\mathcal{X}}^2)}\end{aligned} \tag{22}$$

†† In Appendix E, we prove finiteness of the fundamental SAQFT integrals.

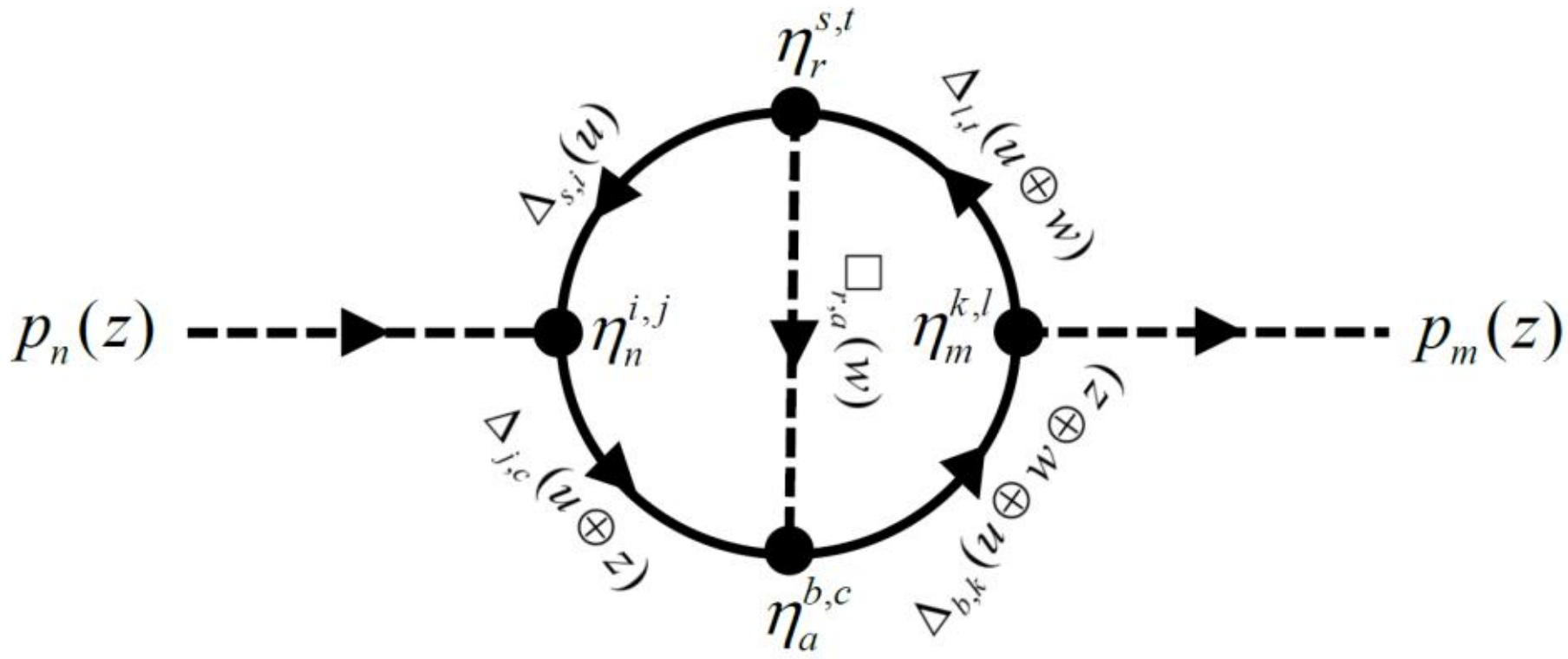

**Fig. 2**: One of the Feynman diagrams contributing to the second order (two-loop) correction to the scalar propagator (i.e., self-energy) in the model $\mathscr{L}_I = \eta \bullet \left[\Phi(\bar{\mathcal{X}}\mathcal{X})\right]$.

In the following section, we evaluate the double integral (21) for a given range of the scalar energy and a fixed set of polynomial degrees $\{r,i,j,k,l\}$ and show that the value of the integral diminishes as the indices become large.

Finally, we consider the first order correction to the interaction vertex in this model. The associated Feynman diagrams are many (in fact, there are six topologically distinct diagrams). We consider here one of these diagrams, which is shown below as Figure 3 where the spectral parameters are related by the energy conservation as $y = x \oplus z$. The bare vertex $\left\langle p_r(z) \middle| q_j(x) q_k(y) \right\rangle_1^0$ is just the coupling tensor element $\eta_r^{j,k}$ independent of the energy. On the other hand, the value of the amputated diagram of Figure 3 reads as follows:

$$\left\langle p_r(z) \middle| q_j(x) q_k(y) \right\rangle_3^1 = \sum_{i,l,n,m,s,t=0}^{\infty} \eta_n^{i,j} \eta_m^{k,l} \eta_r^{s,t} \int \square_{n,m}(u) \Delta_{s,i}(u \oplus x) \Delta_{l,t}(u \oplus y)\, du\,. \tag{23}$$

For monochrome propagation ($n = m$, $s = i$, and $t = l$), this expression simplifies to read:

$$\left\langle p_r(z) \middle| q_j(x) q_k(y) \right\rangle_3^1 = \sum_{i,l,n=0}^{\infty} \eta_n^{i,j} \eta_n^{k,l} \eta_r^{i,l} \int \rho(u)\omega(u \oplus x)\omega(u \oplus y) p_n^2(u) q_i^2(u \oplus x) q_l^2(u \oplus y)\, du\,. \tag{24}$$

In the following section, we show that the value of this *x* and *y* dependent integral falls within the range [0,1] and diminishes rapidly as the polynomial degrees increase.

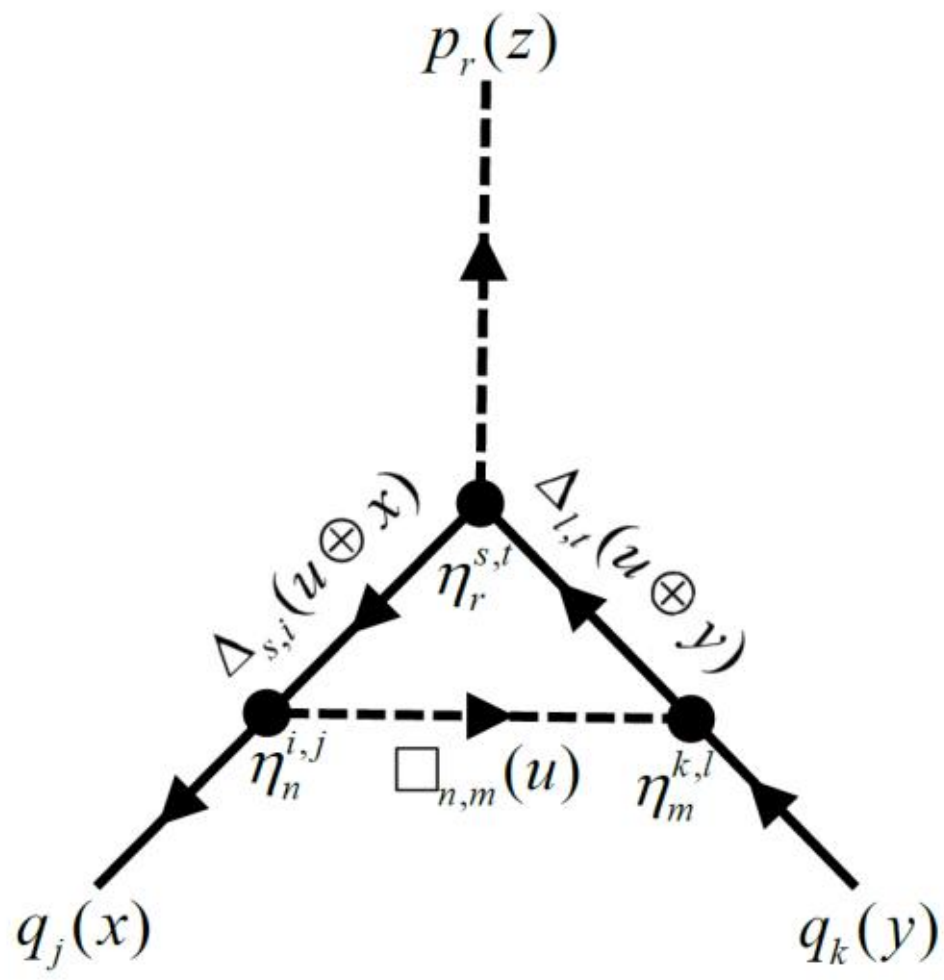

**Fig. 3**: One of the Feynman diagrams contributing to the first order (one-loop) correction to the interaction vertex $\langle p_r(z)|q_j(x)q_k(y)\rangle$ in the model $\mathscr{L}_I = \eta \bullet \left[\Phi(\bar{\mathcal{X}}\mathcal{X})\right]$.

# 4. Numerical results

In this section, we evaluate the sample transition amplitudes represented above by the Feynman diagrams of Figures 1-3. The results demonstrate finiteness of closed-loop integrals for this model in our proposed QFT. For the purpose of calculation, we choose the model parameters as follows. We take a massless scalar ($M_\Phi = 0$) where the associated spectral polynomials $p_n(z)$ is the odd Hermite polynomial $\frac{1}{\sqrt{z}} H_{2n+1}\left(\sqrt{z}\right)$. On the other hand, the spectral polynomial $q_n(z)$ associated with the massive spinor is taken as the Laguerre polynomial $L_n^\nu(z)$. As stated in Section 2, we require that $\nu > 0$ so that the corresponding weight function satisfies the physical constraint $\omega(0) = 0$. For the same reason, we have chosen the odd rather than even Hermite polynomial for the scalar particle so that the corresponding weight function vanishes at zero energy, $\rho(0) = 0$. Now, it is well-known that the odd/even Hermite polynomials could be expressed in terms of the Laguerre polynomials [see Eq. (A3) in Appendix A]. Using that relation, we obtain the following spectral polynomial, its normalized weight function and recursion coefficients associated with the scalar:

$$p_n(z) = (-1)^n \sqrt{n!/\left(\tfrac{3}{2}\right)_n}\, L_n^{1/2}(z), \qquad \rho(z) = \left(2\sqrt{z/\pi}\right) e^{-z}, \tag{25a}$$

$$\alpha_n = 2n + \tfrac{3}{2}, \qquad \beta_n = \sqrt{(n+1)\left(n+\tfrac{3}{2}\right)}, \tag{25b}$$

where $\lambda^2 z = E^2$ and $(x)_n = \Gamma(n+x)/\Gamma(x) = x(x+1)(x+2)\ldots(x+n-1)$ is the Pochhammer symbol (rising/upper factorial). On the other hand, for the spinor they read:

$$q_n(z) = \sqrt{n!/(\nu+1)_n}\, L_n^\nu(z), \qquad \omega(z) = z^\nu e^{-z}/\Gamma(\nu+1), \tag{26a}$$

$$c_n = \sqrt{n+\nu}, \qquad d_n = -\sqrt{n}, \tag{26b}$$

$$\hat{\alpha}_n = c_n^2 + d_{n+1}^2 = 2n + \nu + 1, \qquad \hat{\beta}_n = c_{n+1}\, d_{n+1} = -\sqrt{(n+1)(n+\nu+1)}\,, \tag{26c}$$

where $\lambda^2 z = E^2 - M_{\mathcal{X}}^2$. In the calculation, we take $M_{\mathcal{X}} = \lambda$ and $\nu = 3/2$. Moreover, the elements of the coupling tensor are written in terms of the recursion coefficients (25b) and (26c) as follows

$$\eta_n^{i,j} = \frac{\kappa}{\sqrt{\alpha_n \beta_n}} \left( \frac{1}{\hat{\alpha}_i \hat{\beta}_j} - \frac{1}{\hat{\alpha}_j \hat{\beta}_i} \right) \tag{27}$$

where $\kappa$ is a dimensionless coupling parameter.

For a given set of indices $\{i,j,k,l\}$, we evaluate the integral in (20) that reads

$$I_{i,k,j,l}^{q,q}(z) = \int_0^\infty \omega(u)\omega(u \oplus z) q_i(u) q_k(u) q_j(u \oplus z) q_l(u \oplus z)\, du\,, \tag{28}$$

with $u \oplus z = u + z + 2\sqrt{z(u+1)}$. To evaluate the integral, we use Gauss quadrature integral approximation (see, for example, Ref. [14]). In such calculation, we start by computing the eigenvalues and normalized eigenvectors of an $\mathcal{M} \times \mathcal{M}$ tridiagonal symmetric matrix $J$ whose diagonal elements are $J_{n,n} = \hat{\alpha}_n$ and off-diagonal elements are $J_{n,n+1} = J_{n+1,n} = \hat{\beta}_n$, where $\mathcal{M}$ is the order of the quadrature such that $\mathcal{M} > \max(i,j,k,l)$ and $\{\hat{\alpha}_n, \hat{\beta}_n\}$ are those given by (26c). If we call such eigenvalues $\{\lambda_r\}_{r=0}^{\mathcal{M}-1}$ and the corresponding normalized eigenvectors $\{\Lambda_{s,r}\}_{s=0}^{\mathcal{M}-1}$, then the integral (28) is approximated as follows

$$I_{i,k,j,l}^{q,q}(z) \approx \left( \Lambda W(z)_{j,l} \Lambda^T \right)_{i,k}, \tag{29a}$$

where $W(z)_{j,l}$ is an $\mathcal{M} \times \mathcal{M}$ diagonal matrix whose elements are

$$\left[ W(z)_{j,l} \right]_{r,s} = \delta_{r,s} \left[ \omega(\lambda_r \oplus z) q_j(\lambda_r \oplus z) q_l(\lambda_r \oplus z) \right]. \tag{29b}$$

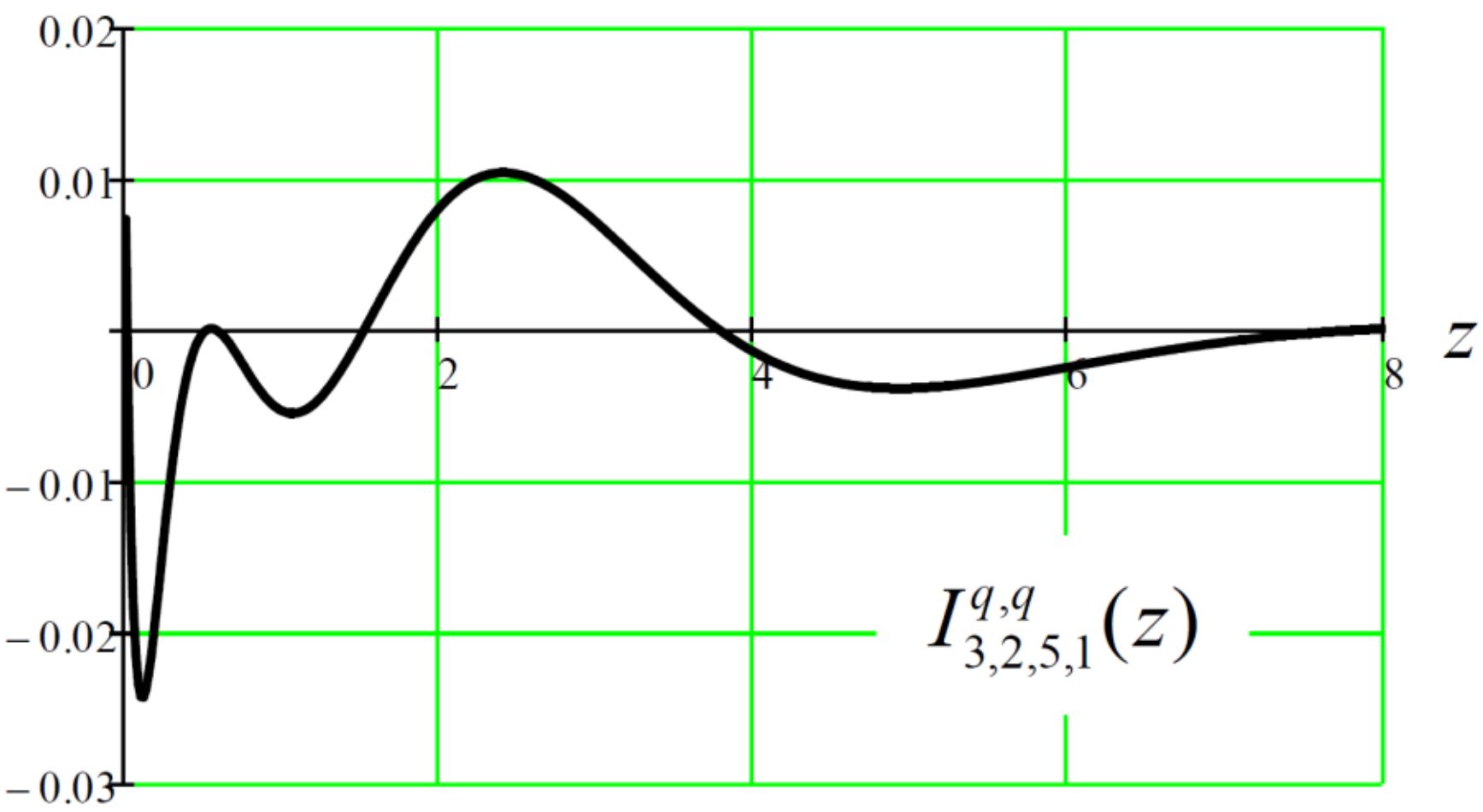

**Fig. 4**: Gauss quadrature approximation of the integral (28) with a quadrature order $\mathcal{M} = 100$.

Figure 4 shows the result of such an evaluation for a given range of the scalar energy $E=\sqrt{z}$. Figure 5 is a plot of $I^{q,q}_{i=k,j=l}(z)=\zeta^{q,q}_{i,j}(z)$ in colors corresponding to monochrome propagation superimposed by several polychrome propagation $I^{q,q}_{i\neq k,j\neq l}(z)$ in black. This figure shows that monochrome propagation is (as expected) positive definite boosting the value of the sum in Eq. (20) whereas polychrome propagation results in cancellations that diminishes the value of the sum. It is also evident from these figures that the absolute value of the integral (28) is less than one.

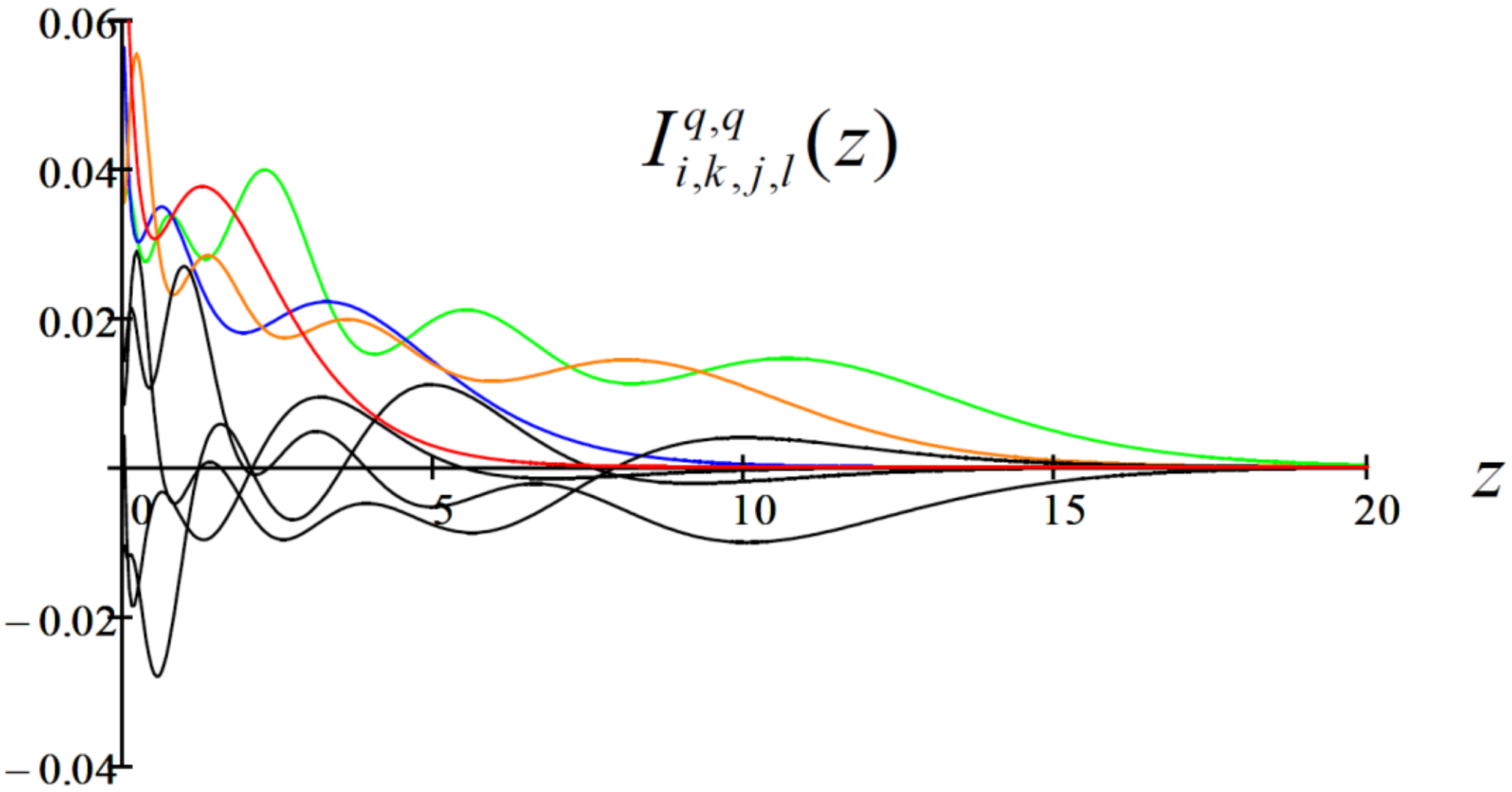

**Fig. 5**: Plot of the integral (28) corresponding to monochrome propagation as $\zeta^{q,q}_{2,1}(z)$ in red, $\zeta^{q,q}_{2,5}(z)$ in green, $\zeta^{q,q}_{3,2}(z)$ in blue, and $\zeta^{q,q}_{3,4}(z)$ in brown. Plots in black correspond to polychrome propagation for $I^{q,q}_{3,2,5,4}(z)$, $I^{q,q}_{2,3,4,1}(z)$, $I^{q,q}_{2,4,5,3}(z)$, and $I^{q,q}_{2,1,2,5}(z)$.

To demonstrate convergence of the series in the calculation of transition amplitudes, we evaluate the correction to the scalar propagator to first order given by Eq. (19) as a function of the number of terms *N* in the sum:

$$S^N(z):=\sum_{i,j,k,l,n',m'=0}^{N}\eta^{i,j}_{n'}\eta^{k,l}_{m'}\Box_{n',m'}(z)I^{q,q}_{i,k,j,l}(z). \tag{30}$$

Figure 6 illustrates convergence of the series $S^N(z)$ as *N* increases where we took $\kappa=1$. Figure 7 is a reproduction of Figure 6 after imposing monochrome propagation where $S^N(z)=\sum_{i,j,n',m'=0}^{N}\eta^{i,j}_{n'}\eta^{i,j}_{m'}\Box_{n',m'}(z)\zeta^{q,q}_{i,j}(z)$. The two figures clearly show that monochrome propagation improves convergence and enhances the value of the transition amplitude. The scalar propagator (19) could then be written to first order as follows:

$$\langle p_m(z)|p_n(z)\rangle\approx\Box_{n,m}(z)\left[1+S^N(z)\right]:=\Box_{n,m}(z)\left[1+\kappa^2 s_\nu(z)\right], \tag{31}$$

where $\kappa^2 s_\nu(z)$ could be interpreted as the square of the energy dependent coupling parameter of the model (to first order).

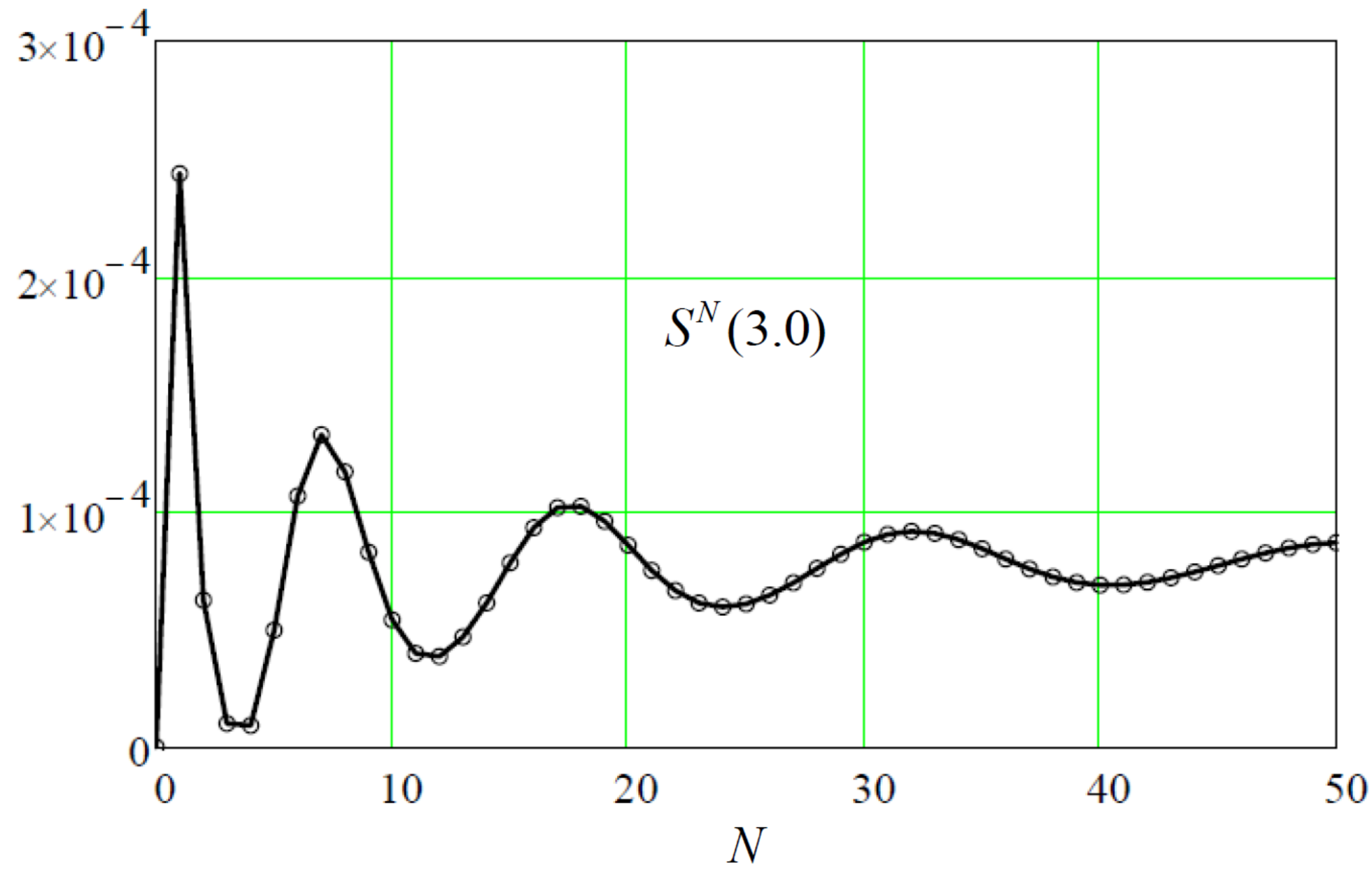

**Fig. 6**: The partial sum $S^N(z)$ defined by Eq. (30) for $\kappa = 1$ and $z = 3.0$. The elements of the coupling tensor in this model are given by Eq. (27).

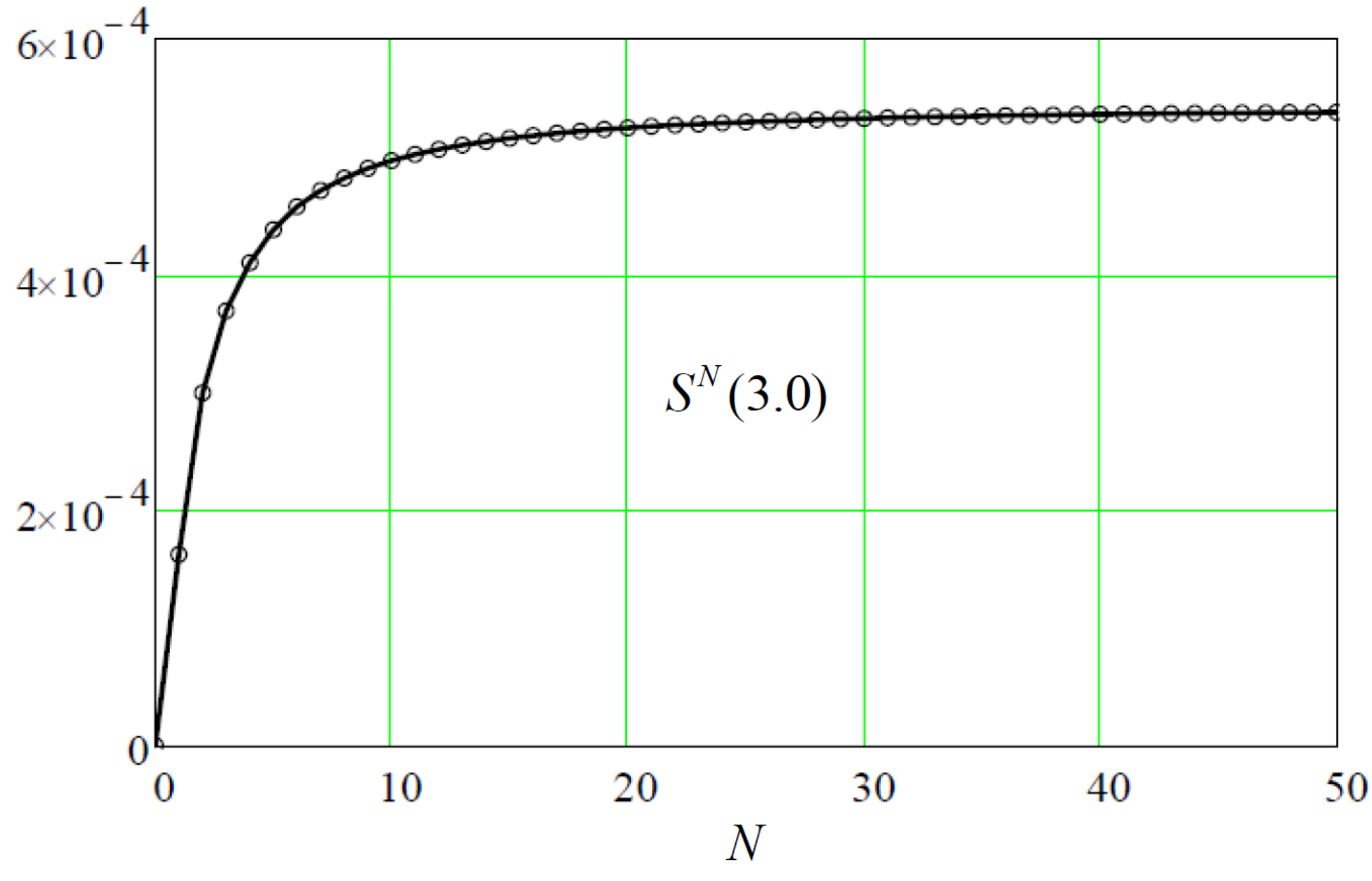

**Fig. 7**: Reproduction of Figure 6 after imposing monochrome propagation.

Table 1 is a list of $s_\nu(z)$ for monochrome propagation and for several values of the energy $E = \sqrt{z}$ and spinor parameter $\nu$. Note that $s_\nu(0) = 0$ since $\square_{n,m}(0) = 0$ and $I_{i,k,j,l}^{q,q}(z) \in [-1,+1]$.

**Table 1**: List of $s_\nu(z)$ for monochrome propagation and for several values of the energy $E=\sqrt{z}$ and spinor parameter $\nu$. In the formula (30), we took $N=100$.

| | $s_0(z)\times10^3$ | $s_{\frac{1}{2}}(z)\times10^3$ | $s_1(z)\times10^3$ | $s_{\frac{3}{2}}(z)\times10^3$ | $s_2(z)\times10^3$ |
|---|---|---|---|---|---|
| $z=0.5$ | 10.335 | 3.937 | 1.986 | 1.158 | 0.739 |
| $z=1.0$ | 8.458 | 3.230 | 1.638 | 0.968 | 0.628 |
| $z=2.0$ | 4.408 | 1.733 | 0.895 | 0.537 | 0.354 |
| $z=3.0$ | 2.281 | 0.920 | 0.483 | 0.294 | 0.196 |
| $z=4.0$ | 1.249 | 0.513 | 0.273 | 0.168 | 0.113 |
| $z=5.0$ | 0.723 | 0.301 | 0.162 | 0.100 | 0.068 |

The Table indicates that the value of $s_\nu(z)$ decreases as the energy or parameter $\nu$ increase. If we repeat the same self-energy calculation but for the spinor propagator, we obtain the following result to first order

$$\langle q_m(z)|q_n(z)\rangle \approx \Delta_{n,m}(z)\times \left[1+\sum_{i,j,k,l,n',m'=0}^{N}\eta_i^{n',k}\eta_j^{l,m'}\Delta_{n',m'}(z)\int_0^\infty \square_{i,j}(u)\Delta_{k,l}(u\oplus z)\,du\right]=\Delta_{n,m}(z)\left[1+S^N(z)\right] \tag{32}$$

Figure 8 is a reproduction of Figures 4 and 5 for the truncated sum in (32) for the spinor.

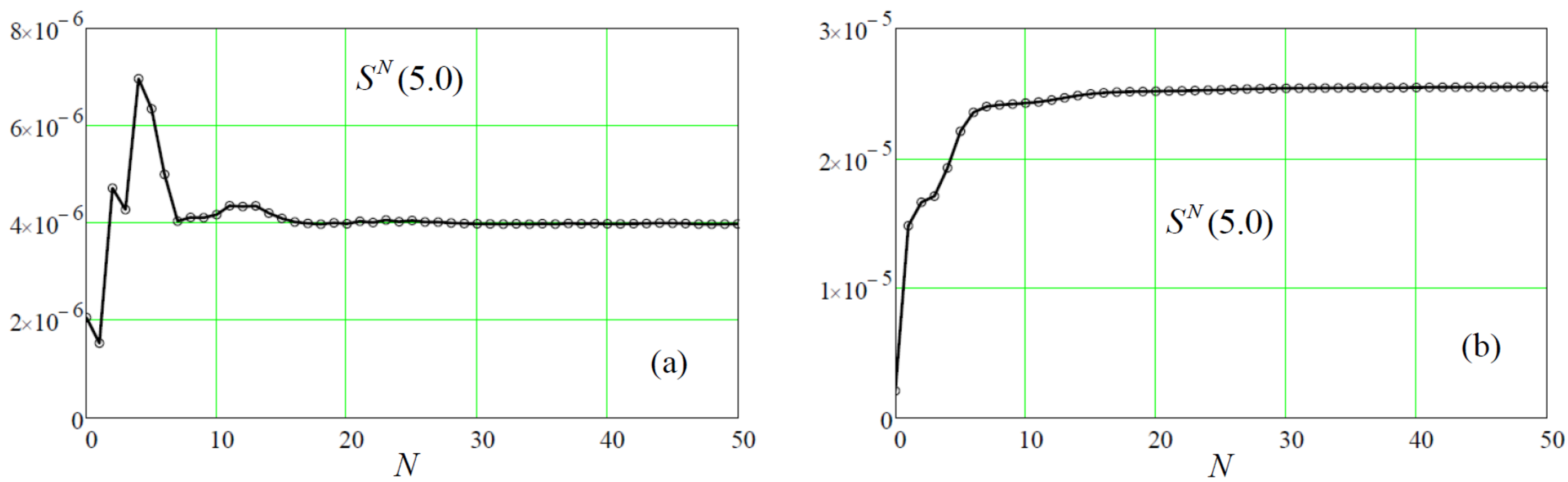

**Fig. 8**: Reproduction of Figures 4 and 5 for the spinor particle but for $z=5.0$: (a) polychrome propagation, (b) monochrome propagation.

Next, we evaluate the double integral associated with the diagram of Figure 2 for monochrome propagation and given in Eq. (21) that reads

$$
\begin{aligned}
\zeta^{p,q,q,q,q}_{r,i,j,l,k}(z) = \int_0^\infty \int_0^\infty & \rho(w)\omega(u)\omega(u\oplus z)\omega(u\oplus w)\omega(u\oplus w\oplus z) \\
& \times p_r^2(w) q_i^2(u) q_j^2(u\oplus z) q_l^2(u\oplus w) q_k^2(u\oplus w\oplus z)\, du\, dw
\end{aligned}
\tag{33}
$$

where $u\oplus z = u+z+2\sqrt{z(u+1)}$, $u\oplus w = u+w+2\sqrt{w(u+1)}$ and $u\oplus w\oplus z = u+w+z+2\sqrt{zw}+2\sqrt{z(u+1)}+2\sqrt{w(u+1)}$. Figure 9 is a plot of $\zeta^{p,q,q,q,q}_{r,i,j,l,k}(z)$ for a given set of indices and range of scalar energy. The figure is a superposition of three evaluations of the integral using an increasing order of Gauss quadrature: $\mathcal{M}=15$ (dashed red), $\mathcal{M}=20$ (dotted blue), and $\mathcal{M}=30$ (solid black).

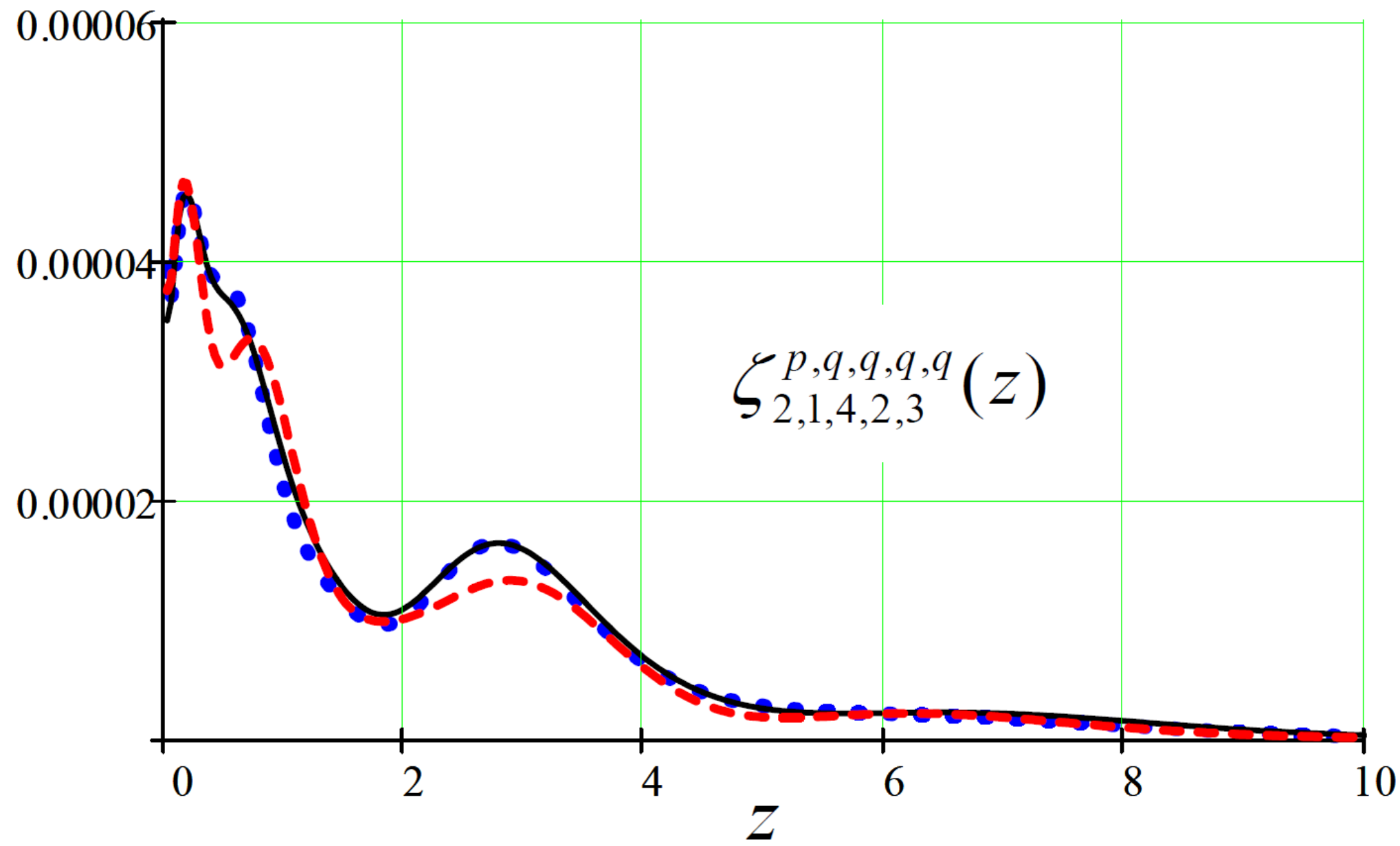

**Fig. 9**: Plot of the integral (33) demonstrating convergence by increasing the Gauss quadrature order from 15 (dashed red), to 20 (dotted blue), to 30 (solid black).

Figure 10 is a reproduction of Figure 9 (with a Gauss quadrature order $\mathcal{M}=50$) but for large values of the indices (10 times those of Figure 9). The figure demonstrates diminishing values for the integral (of the order of 200 times less). For this latter calculation, we used the large degree asymptotics of the Laguerre polynomials to write

$$
\lim_{n\to\infty} q_n(z) = \frac{\sqrt{\Gamma(\nu+1)/\pi}}{n^{\frac{1}{4}} z^{\frac{\nu}{2}+\frac{1}{4}} e^{-z/2}} \sin\left(2\sqrt{nz}-\tfrac{\pi\nu}{2}+\tfrac{\pi}{4}\right) + O\left(n^{-\frac{3}{4}}\right),
\tag{34a}
$$

$$
\lim_{n\to\infty} p_n(z) = \frac{(-1)^n e^{z/2}}{(n\pi)^{\frac{1}{4}}} \cos\left(2\sqrt{nz}\right) + O\left(n^{-\frac{3}{4}}\right),
\tag{34b}
$$

where we have also used $\lim\limits_{n\to\infty}\left[\Gamma(n+x)/\Gamma(n+y)\right] = n^{x-y}$.

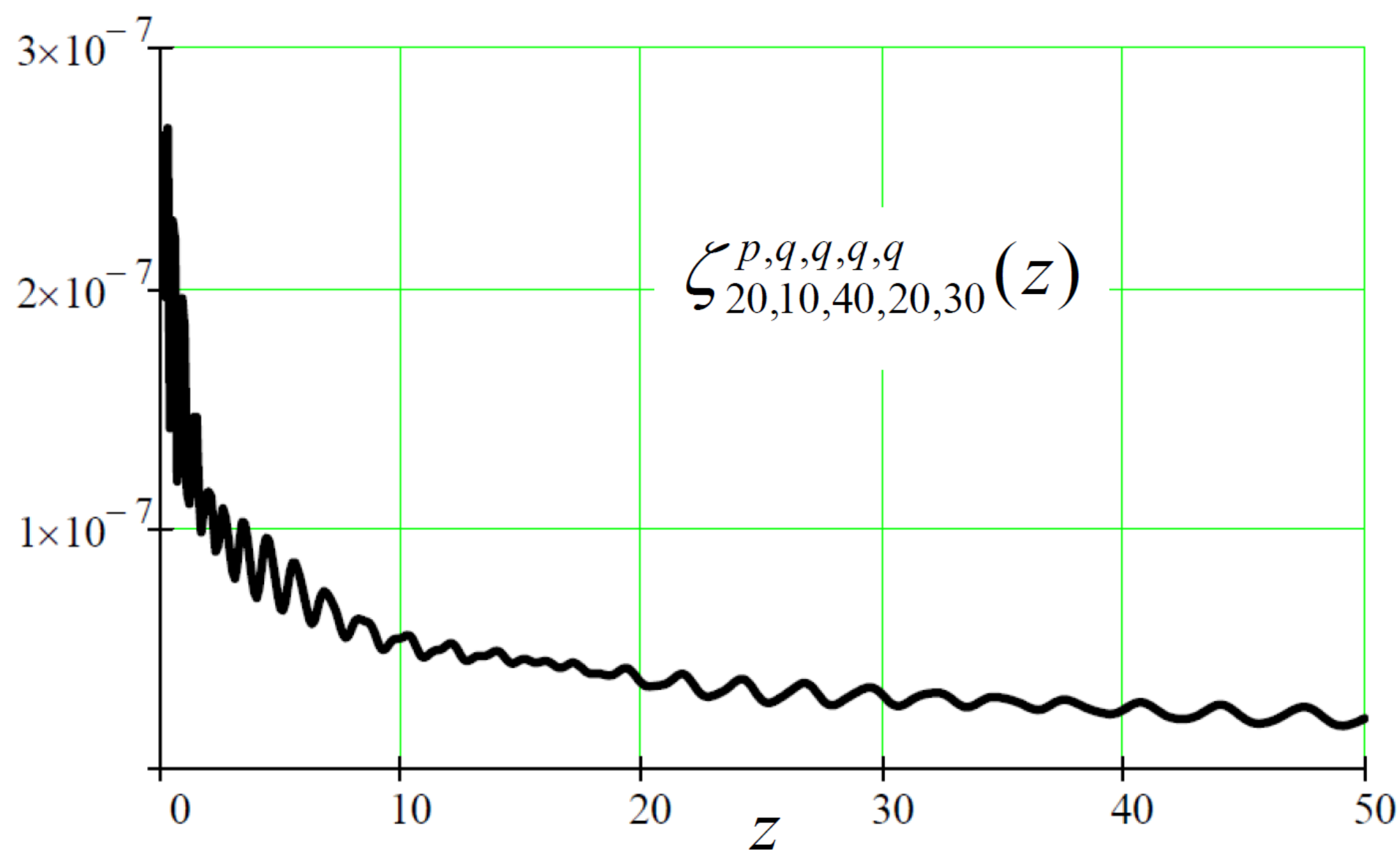

**Fig. 10**: Reproduction of Figure 9 (with a Gauss quadrature order of 50) but for 10 times the values of the indices demonstrating diminishing values (of the order of 200 times less) where we used the asymptotic formula (34) for the spectral polynomials.

Finally, we evaluate the integral in the vertex diagram of Figure 3 for monochrome propagation as given in Eq. (24) that reads

$$\zeta^{p,q,q}_{n,i,l}(x,y)=\int_0^\infty \rho(u)\omega(u\oplus x)\omega(u\oplus y)p_n^2(u)q_i^2(u\oplus x)q_l^2(u\oplus y)du\,, \tag{35}$$

where $u\oplus x=u+x+2\sqrt{u(x+1)}$ and $u\oplus y=u+y+2\sqrt{u(y+1)}$. Figure 11 is a 3D plot of $\zeta^{p,q,q}_{n,i,l}(x,y)$ for the indices $\{n,i,l\}=\{3,2,5\}$ and spinor particle energy ranges $x\in[0,15]$ and $y\in[0,30]$.

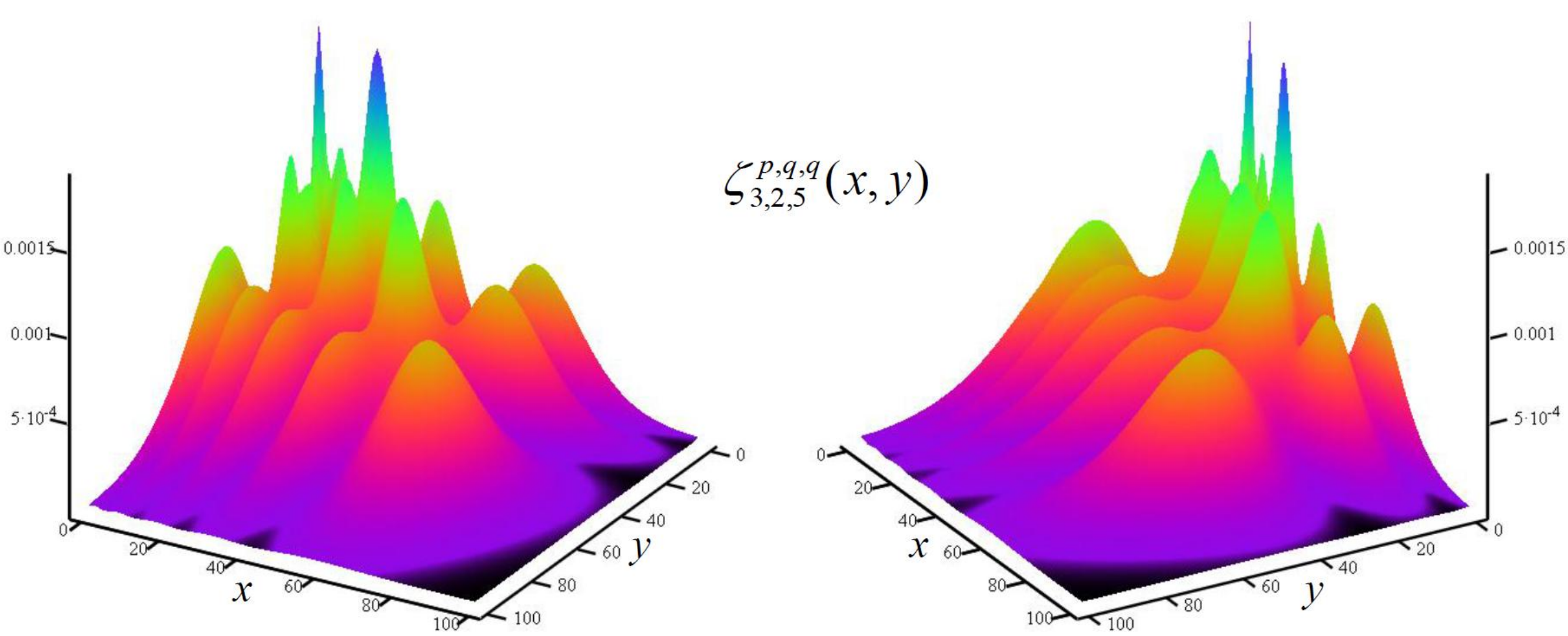

**Fig. 11**: Two-dimensional plot of the integral (35) viewed from two different angles. The *x*-axis scale is 0.15 while the *y*-axis scale is 0.30.

## 5. Conclusion

We presented in this introductory study an algebraic version of QFT that utilizes orthogonal polynomials and where we adopt a spectral decomposition of the quantum fields in the energy rather than linear momentum space. The quantum fields in this theory are created, annihilated, propagated and coupled on the level of individual spectral modes of the field rather than in totality as in conventional QFT. The differential wave equation for the free quantum field in conventional QFT is replaced by an algebraic three-term recursion relation for the associated spectral polynomials. The proposed theory is, in fact, equivalent to the structureless sector of SAQFT that was introduced recently in [13]. We presented several examples in a typical interaction model illustrating the remarkable property of the theory that it is finite; eliminating the need for renormalization altogether. It remains to be seen whether this finiteness property prevails in other physically relevant models within this spectral QFT. In Appendices B, C, and D, we give a brief description of how to construct the spinor, massless and massive vector fields in this theory. Moreover, in Appendix E, we give a proof of the finiteness of closed loop integrals in the Feynman diagrams of this spectral theory.

Finally, we like to state that the work presented here should be considered the start of the development of a robust quantum field theory of elementary particles that does not require renormalization. We do acknowledge that the formulation of the theory as presented in this work needs further development and/or improvement. For example, developing a more efficient calculation of the integrals in the Feynman diagrams (other than the Gauss quadrature that we used above) and making improvements in doing sums in the transition amplitudes. Moreover, additional investigation is required to determine whether our conclusions and findings hold in more general settings. However, in a single introductory article, it is not possible to address all relevant issues of the theory that have been investigated in a large volume of research for over a century of conventional QFT. Nonetheless, a top priority could be to reproduce well-known and/or established results (e.g., the anomalous magnetic moment of the electron in QED). If Finite Spectral QFT can be fully reconciled with established findings in nuclear and particle physics, it would mean that the singularities that we see in conventional QFT aren't fundamental properties of nature, but rather artifacts of using the wrong representations or incorrect mathematical tools. On the other hand, the findings of our present study indicate that such finite theory does exist indeed and that it is a viable alternative to the conventional formulation of QFT.

## Acknowledgement:

I am grateful to Dr. A. Bahaoui and the theoretical physics group lead by Prof. A. Jellal at Chouaib Doukkali University in El Jadida, Morocco for an independent verification of our result shown in Figure 6.

## Appendix A: Equivalence of the linear momentum representation in conventional QFT to the energy representation in spectral QFT

In conventional QFT, the positive energy component of the free quantum field associated with a scalar particle in 1+1 space-time is written as the following continuous Fourier expansion in the linear momentum *k*-space [5-7]

$$\Phi(t,x)=\int\frac{dk}{\sqrt{2E}}e^{-\mathrm{i}Et}\,\frac{e^{\mathrm{i}kx}}{\sqrt{2\pi}}\mathsf{a}(k)\,, \tag{A1}$$

where the creation/annihilation field operators satisfy the commutation algebra $[\mathsf{a}(k),\mathsf{a}^{\dagger}(k')]=\mathsf{a}(k)\mathsf{a}^{\dagger}(k')-\mathsf{a}^{\dagger}(k')\mathsf{a}(k)=\delta(k-k')$. The free Kelin-Gordon wave equation $\left(\partial_x^2-\partial_t^2\right)\Phi(t,x)=M^2\Phi(t,x)$ gives the energy-momentum relation for a free particle as $E^2=k^2+M^2$. Particle excitations are created from the vacuum as the continuous wavefunction $\chi(t,x)=e^{-\mathrm{i}Et}e^{\mathrm{i}kx}$.

Using Mehler's formula (See problem 23 on page 380 of Ref. [8]), the oscillatory function $\frac{e^{\mathrm{i}kx}}{\sqrt{2\pi}}$ (plane wave) has the following point-wise convergent expansion

$$\frac{e^{\mathrm{i}kx}}{\sqrt{2\pi}}=\frac{1}{\sqrt{\pi}}e^{-k^2/2\lambda^2}e^{-\lambda^2x^2/2}\sum_{n=0}^{\infty}\frac{\mathrm{i}^n}{2^n\,n!}H_n(k/\lambda)H_n(\lambda x)\,, \tag{A2}$$

where $H_n(y)$ is the Hermite polynomial of degree $n$ in $y$, and $\lambda$ is an arbitrary scale parameter with the dimension of mass/energy. The even and odd Hermite polynomials are related to the Laguerre polynomials as follows

$$H_{2n}(y)=(-1)^n2^{2n}n!L_n^{-1/2}(y^2)\,,\qquad H_{2n+1}(y)=(-1)^n2^{2n+1}n!\,yL_n^{+1/2}(y^2)\,, \tag{A3}$$

Thus, we can rewrite Eq. (A2) as follows:

$$\frac{e^{\mathrm{i}kx}}{\sqrt{2\pi}}=\frac{1}{\sqrt{\pi}}e^{-z/2}e^{-\lambda^2x^2/2}\left[\sum_{n=0}^{\infty}\frac{n!}{2n!}L_n^{-1/2}(z)H_{2n}(\lambda x)+\mathrm{i}\sqrt{z}\sum_{n=0}^{\infty}\frac{n!}{(2n+1)!}L_n^{+1/2}(z)H_{2n+1}(\lambda x)\right], \tag{A4}$$

where we wrote $k^2/\lambda^2=z$. The orthonormal version of the Laguerre polynomials are $p_n^{\pm}(z):=\sqrt{\frac{n!\Gamma(1\pm\frac{1}{2})}{\Gamma(n+1\pm\frac{1}{2})}}L_n^{\pm1/2}(z)$ and the corresponding normalized weight functions are $\rho^{\pm}(z)=z^{\pm\frac{1}{2}}e^{-z}/\Gamma(1\pm\frac{1}{2})$ whereas the recursion coefficients read

$$\alpha_n^{\pm}=2n\pm\tfrac{1}{2}+1\,,\qquad \beta_n^{\pm}=-\sqrt{(n+1)\left(n\pm\tfrac{1}{2}+1\right)}\,. \tag{A5}$$

Therefore, we can rewrite Eq. (A4) as

$$\frac{e^{\mathrm{i}kx}}{\sqrt{2\pi}}=\left(z/\pi\right)^{1/4}e^{-\lambda^2x^2/2}\left[\sqrt{\rho^-(z)}\sum_{n=0}^{\infty}p_n^-(z)\frac{H_{2n}(\lambda x)}{\sqrt{2^{2n}(2n)!}}+\mathrm{i}\sqrt{\rho^+(z)}\sum_{n=0}^{\infty}p_n^+(z)\frac{H_{2n+1}(\lambda x)}{\sqrt{2^{2n+1}(2n+1)!}}\right], \tag{A6}$$

where we have used the identities: $\Gamma\left(n+\frac{1}{2}\right)=\frac{(2n)!\sqrt{\pi}}{2^{2n}n!}$, $\Gamma(z+1)=z\Gamma(z)$, and $\Gamma\left(\frac{1}{2}\right)=\sqrt{\pi}$. Equation (A6) could also be written as

$$\frac{e^{\mathrm{i}kx}}{\sqrt{2\pi}}=\sqrt{\frac{k}{\lambda}}\left[\sqrt{\rho^-(z)}\sum_{n=0}^{\infty}p_n^-(z)\phi_n^-(x)+\mathrm{i}\sqrt{\rho^+(z)}\sum_{n=0}^{\infty}p_n^+(z)\phi_n^+(x)\right], \tag{A7}$$

where $\phi_n^\pm(x) = \left(e^{-\lambda^2 x^2/2}\big/\pi^{1/4}\right)\left[H_{m_\pm}(\lambda x)\big/\sqrt{2^{m_\pm} m_\pm!}\right]$ and $m_\pm = 2n + \frac{1\pm1}{2}$. Using the orthogonality of the Hermite polynomials, $\int_{-\infty}^{+\infty} e^{-y^2} H_n(y) H_m(y) dy = \sqrt{\pi}\, 2^n n!\, \delta_{n,m}$, one can show that $\left\langle \phi_n^+(x) \middle| \phi_{n'}^-(x) \right\rangle = 0$ and $\left\langle \phi_n^\pm(x) \middle| \bar{\phi}_{n'}^\pm(x) \right\rangle = \delta_{n,n'}$ where $\bar{\phi}_n^\pm(x) = \lambda \phi_n^\pm(x)$. Using the differential equation of the Hermite polynomials, $H_n''(y) - 2y H_n'(y) + 2n H_n(y) = 0$, and their recursion relation, $y H_n(y) = n H_{n-1}(y) + \frac{1}{2} H_{n+1}(y)$, on can show that

$$-\partial_x^2 \phi_n^\pm(x) = \lambda^2 \left[ \alpha_n^\pm \phi_n^\pm(x) + \beta_{n-1}^\pm \phi_{n-1}^\pm(x) + \beta_n^\pm \phi_{n+1}^\pm(x) \right]. \tag{A8}$$

Substituting (A7) in (A1) and using the on-shell relation for free fields $k dk = E dE$, we can write

$$\Phi(t,x) = \int dE \sqrt{2E}\, e^{-\mathrm{i}Et} \sum_{n=0}^{\infty} \left[ \tfrac{1}{\sqrt{2}} a_n^-(E) \phi_n^-(x) + \tfrac{\mathrm{i}}{\sqrt{2}} a_n^+(E) \phi_n^+(x) \right], \tag{A9}$$

where we wrote $a_n^\pm(E) = \sqrt{\rho^\pm(z)/2\lambda k}\; p_n^\pm(z) \mathsf{a}(k)$. Accordingly, particle excitations are created from the vacuum as the discrete wavefunctions $\chi_n^\pm(t,x) = e^{-\mathrm{i}Et} \phi_n^\pm(x)$. Moreover, the following becomes a map between the field operators in the two representations: $\mathsf{a}(k) \mapsto a_0(E)\sqrt{2\lambda k/\rho(z)}$.

In 3+1 Minkowski space-time, the positive energy component of the free quantum field associated with a scalar particle in conventional QFT is written as the following Fourier expansion in the linear momentum $\vec{k}$ -space:

$$\Phi(t,\vec{r}) = \frac{1}{(2\pi)^{3/2}} \int \frac{d^3\vec{k}}{\sqrt{2E}} e^{-\mathrm{i}Et} e^{\mathrm{i}\vec{k}\cdot\vec{r}} \mathsf{a}(\vec{k}). \tag{A10}$$

In spherical coordinates, $\vec{r} = (r, \vartheta, \varphi)$ and we set the *z*-axis along $\vec{k}$ making $e^{\mathrm{i}\vec{k}\cdot\vec{r}} = e^{\mathrm{i}\kappa r \cos\vartheta}$ where $\kappa = |\vec{k}|$. The creation/annihilation field operators satisfy the commutation algebra $[\mathsf{a}(\vec{k}), \mathsf{a}^\dagger(\vec{k}')] = \mathsf{a}(\vec{k})\mathsf{a}^\dagger(\vec{k}') - \mathsf{a}^\dagger(\vec{k}')\mathsf{a}(\vec{k}) = \delta^3(\vec{k} - \vec{k}')$. The free Kelin-Gordon wave equation

$$\left( \vec{\nabla}^2 - \partial_t^2 \right) \Phi(t,\vec{r}) = M^2 \Phi(t,\vec{r}), \tag{A11}$$

gives the energy-momentum relation for the free particle as $E^2 = \kappa^2 + M^2$. An alternative to Mehler's formula (A2) that is more suitable for this space-time reads as follows (see, for example, Eq. (4.8.3) in Ref. [11]):

$$e^{\mathrm{i}kx} = 2^\nu \Gamma(\nu) (\lambda x)^{-\nu} \sum_{n=0}^{\infty} \mathrm{i}^n (n+\nu) C_n^\nu(k/\lambda) J_{n+\nu}(\lambda x), \tag{A12}$$

where $C_n^\nu(y)$ is the Gegenbauer (ultra-spherical) polynomial, $J_{n+\nu}(y)$ is the Bessel function of the first kind and $\nu$ is an arbitrary dimensionless positive parameter. Since $e^{\mathrm{i}\vec{k}\cdot\vec{r}} = e^{\mathrm{i}\kappa r\cos\vartheta}$, then if we take $x = r$, $k = \kappa\cos\vartheta$, and $\lambda = \kappa$, Eq. (A12) gives the following expansion for the plane wave $e^{\mathrm{i}\vec{k}\cdot\vec{r}}$

$$e^{\mathrm{i}\vec{k}\cdot\vec{r}}=e^{\mathrm{i}\kappa r\cos\vartheta}=2^{\nu}\Gamma(\nu)\left(\kappa r\right)^{-\nu}\sum_{\ell=0}^{\infty}\mathrm{i}^{\ell}(\ell+\nu)C_{\ell}^{\nu}(\cos\vartheta)J_{\ell+\nu}(\kappa r)\,, \tag{A13}$$

where we have made the suggestive 3-dimensional index replacement $n\mapsto\ell$ with $\ell$ being the angular momentum quantum number. On the other hand, it is known that the Bessel function has the following series representation [8,9]

$$J_{\mu}(\kappa r)=2(\kappa r)^{\mu}e^{-\lambda^2r^2/2}e^{-\kappa^2/2\lambda^2}\sum_{n=0}^{\infty}\frac{(-1)^n n!}{\Gamma(n+\mu+1)}L_n^{\mu}(\kappa^2/\lambda^2)L_n^{\mu}(\lambda^2r^2)\,. \tag{A14}$$

Therefore, we can rewrite (A13) as the following double expansion

$$e^{\mathrm{i}\vec{k}\cdot\vec{r}}=2^{\nu+1}\Gamma(\nu)e^{-y/2}e^{-z/2}\sum_{n,\ell=0}^{\infty}\frac{\mathrm{i}^{\ell}(\ell+\nu)(-1)^n n!}{\Gamma(n+\ell+\nu+1)}C_{\ell}^{\nu}(\cos\vartheta)(yz)^{\ell/2}L_n^{\ell+\nu}(z)L_n^{\ell+\nu}(y)\,, \tag{A15}$$

where $y=\lambda^2r^2$ and $z=\kappa^2/\lambda^2=(E^2-M^2)/\lambda^2$. Now, we define the spectral polynomial as $p_n^{\ell}(z):=(-1)^n\sqrt{n!/\left(\ell+\nu+1\right)_n}\,L_n^{\ell+\nu}(z)$ whose weight function is $\rho^{\ell}(z)=z^{\ell+\nu}e^{-z}/\Gamma(\ell+\nu+1)$ and associated recursion coefficients are:

$$\alpha_n^{\ell}=2n+\ell+\nu+1\,,\qquad\qquad \beta_n^{\ell}=\sqrt{(n+1)\left(n+\ell+\nu+1\right)}\,. \tag{A16}$$

Moreover, we define the spatial functions $\phi_n^{\ell}(\vec{r}):=\chi_n^{\ell}(y)Y_{\ell}(\vartheta)$ where:

$$\chi_n^{\ell}(y):=A_n^{\ell}(\nu)y^{\ell/2}e^{-y/2}L_n^{\ell+\nu}(y)\,, \tag{A17a}$$

$$Y_{\ell}(\vartheta):=\mathrm{i}^{\ell}B_{\ell}(\nu)C_{\ell}^{\nu}(\cos\vartheta)\,, \tag{A17b}$$

with $A_n^{\ell}(\nu)$ and $B_{\ell}(\nu)$ being real normalization constants. The differential equation and recursion relation of the Laguerre polynomial together with the differential equation of the Gegenbauer polynomial can be used to show that $\phi_n^{\ell}(\vec{r})$ satisfies Eq. (3) with the recursion coefficients (A16) if and only if $\nu=\frac{1}{2}$ where in spherical coordinates $\vec{\nabla}^2=\frac{\partial^2}{\partial r^2}+\frac{2}{r}\frac{\partial}{\partial r}+\frac{1}{r^2}L(\vartheta,\varphi)$ and $L(\vartheta,\varphi)=\frac{\partial^2}{\partial\vartheta^2}+\frac{\cos\vartheta}{\sin\vartheta}\frac{\partial}{\partial\vartheta}+\frac{1}{\sin^2\vartheta}\frac{\partial^2}{\partial\varphi^2}$. Another justification for the choice $\nu=\frac{1}{2}$ is given next.

The conjugate function $\bar{\phi}_n^{\ell}(\vec{r})$ is defined in the sense of the orthogonality (6a) with respect to the integration measure $d^3\vec{r}=r^2dr\sin\vartheta d\vartheta d\varphi$. That is, $\left\langle\phi_n^{\ell}(\vec{r})\middle|\bar{\phi}_{n'}^{\ell'}(\vec{r})\right\rangle=\delta_{n,n'}\delta_{\ell,\ell'}$. Using the orthogonality of the Laguerre and Gegenbauer polynomials [8,9], we obtain

$$\bar{\chi}_n^{\ell}(y)=\lambda A_n^{\ell}(\nu)y^{\nu+\frac{\ell-1}{2}}e^{-y/2}L_n^{\ell+\nu}(y)\,, \tag{A18a}$$

$$\bar{Y}_{\ell}(\vartheta):=(-\mathrm{i})^{\ell}B_{\ell}(\nu)\left(\sin\vartheta\right)^{2\nu-1}C_{\ell}^{\nu}(\cos\vartheta)\,. \tag{A18b}$$

Consequently, a proper choice of the arbitrary parameter $\nu$ that makes $\phi_n^{\ell}(\vec{r})$ self-conjugate [i.e., $\bar{\phi}_n^{\ell}(\vec{r})\propto\phi_n^{\ell}(\vec{r})^*$] is $\nu=\frac{1}{2}$ giving the following normalization constants:

$$A_n^\ell(\nu)=\sqrt{2\lambda(n!)/\Gamma\left(n+\ell+\tfrac{3}{2}\right)}\,,\qquad B_\ell(\nu)=\sqrt{\left(\ell+\tfrac{1}{2}\right)/2\pi}\,. \tag{A19}$$

Substituting (A17) with $\nu=\frac{1}{2}$ and (A19) in (A15) results in the following double expansion formula

$$e^{i\vec{k}\cdot\vec{r}}=\frac{2^{3/2}\pi}{\sqrt{\kappa}}\sum_{n,\ell=0}^{\infty}\sqrt{\left(\ell+\tfrac{1}{2}\right)\rho^\ell(z)}\,p_n^\ell(z)\phi_n^\ell(\vec{r})\,. \tag{A20}$$

Inserting this expression in (A10) with $d^3\vec{k}=4\pi\kappa^2 d\kappa$ (since the $z$-axis is taken along $\vec{k}$ ) and using $E^2=\kappa^2+M^2$ to write $\kappa d\kappa=EdE$, we obtain

$$\Phi(t,\vec{r})=\int dE\sqrt{2E}\,e^{-iEt}\left(\sum_{n,\ell=0}^{\infty}\sqrt{2\pi\left(2\ell+1\right)\kappa\,\rho^\ell(z)}\,p_n^\ell(z)\phi_n^\ell(\vec{r})\right)\mathsf{a}(\vec{k})\,. \tag{A21}$$

Writing it in the Spectral QFT representation, $\Phi(t,\vec{r})=\int dE\sqrt{2E}\,e^{-iEt}\sum_{n,\ell=0}^{\infty}a_n^\ell(E)\phi_n^\ell(\vec{r})$ with $a_n^\ell(E)=p_n^\ell(z)a_0^\ell(E)$, we obtain the following operator map $a_0^\ell(E)=\sqrt{2\pi(2\ell+1)\kappa\,\rho^\ell(z)}\,\mathsf{a}(\vec{k})$.

## Appendix B: Dirac spinor in spectral QFT

We provide in this Appendix a brief description of the four-component quantum field $\Psi^{\uparrow\downarrow}(t,\vec{r})$ representing the Dirac spinor with spin up $\uparrow$ and down $\downarrow$ in 3+1 Minkowski space-time. If we adopt the conventional representation of the $4\times 4$ Dirac gamma matrices as $\gamma^0=\begin{pmatrix} \mathrm{I} & 0\\ 0 & -\mathrm{I}\end{pmatrix}$ and $\vec{\gamma}=\begin{pmatrix} 0 & \vec{\sigma}\\ -\vec{\sigma} & 0\end{pmatrix}$, where I is the $2\times 2$ unit matrix and $\{\vec{\sigma}\}$ are the three Pauli spin matrices, then the free Dirac equation for $\Psi^{\uparrow\downarrow}(t,\vec{r})$ reads $\sum_{\mu=0}^{\mu=3}\left(i\gamma^\mu\partial_\mu-M\right)\Psi^{\uparrow\downarrow}(t,\vec{r})=0.$ Multiplying the equation from left by $\gamma^0$, we obtain the following $4\times 4$ matrix wave equation

$$\begin{pmatrix}\left(i\partial_t-M\right)\mathrm{I} & i\vec{\sigma}\cdot\vec{\nabla}\\ i\vec{\sigma}\cdot\vec{\nabla} & \left(i\partial_t+M\right)\mathrm{I}\end{pmatrix}\begin{pmatrix}\Psi_+^{\uparrow\downarrow}(t,\vec{r})\\ \Psi_-^{\uparrow\downarrow}(t,\vec{r})\end{pmatrix}=0\,. \tag{B1}$$

The two components of the quantum spinor field are written as

$$\Psi_\pm^\uparrow(t,\vec{r})=\int dE\sqrt{2E}\,e^{-iEt}\sum_{n=0}^{\infty}b_n^\uparrow(E)\phi_n^\pm(\vec{r})\,, \tag{B2a}$$

$$\Psi_\pm^\downarrow(t,\vec{r})=\int dE\sqrt{2E}\,e^{-iEt}\sum_{n=0}^{\infty}b_n^\downarrow(E)\phi_n^\pm(\vec{r})\,, \tag{B2b}$$

where $b_n^{\uparrow\downarrow}(E)=b_0^{\uparrow\downarrow}(E)p_n^{\uparrow\downarrow}(z)$ making $p_0^{\uparrow\downarrow}(z)=1$ and the spectral parameter $z$ is to be determined. The creation and annihilation operators satisfy the following anti-commutation algebra

$$\left\{b_0^s(E),b_0^{s'\dagger}(E')\right\}=b_0^s(E)b_0^{s'\dagger}(E')+b_0^{s'\dagger}(E')b_0^s(E)=\delta_{s,s'}\,f^s(E)\delta(E-E')\,, \tag{B3}$$

where $s$ and $s'$ stand for the $\uparrow\downarrow$ spins and $f^s(E)$ is a positive definite function to be determined below by canonical quantization. $\left\{\phi_n^\pm(\vec{r})\right\}$ are two-component spatial functions that satisfy the following coupled differential relations

$$-\mathrm{i}\vec{\sigma}\cdot\vec{\nabla}\phi_n^-(\vec{r}) = \lambda\left[c_n\,\phi_n^+(\vec{r}) + d_n\,\phi_{n-1}^+(\vec{r})\right], \tag{B4a}$$

$$-\mathrm{i}\vec{\sigma}\cdot\vec{\nabla}\phi_n^+(\vec{r}) = \lambda\left[c_n\,\phi_n^-(\vec{r}) + d_{n+1}\,\phi_{n+1}^-(\vec{r})\right], \tag{B4b}$$

where $\left\{c_n, d_n\right\}$ are dimensionless constants such that $c_n d_{n+1} > 0$ for all $n$. The first (second) relation above is referred to as the forward (backward) shift operator. Using (B4), the coupled free Dirac equation (B1) becomes the following set of two algebraic difference relations

$$\left(E+M\right)p_n^s(z) = \lambda\left[c_n\,p_n^s(z) + d_n\,p_{n-1}^s(z)\right], \qquad \text{for } E < -M\,. \tag{B5a}$$

$$\left(E-M\right)p_n^s(z) = \lambda\left[c_n\,p_n^s(z) + d_{n+1}\,p_{n+1}^s(z)\right], \qquad \text{for } E > +M\,. \tag{B5b}$$

Multiplying both sides of Eq. (B5a) by $E-M$ for $E > +M$ and using Eq. (B5b), we obtain the following algebraic relation

$$\left(E^2-M^2\right)p_n^s(z) = \lambda^2\left[\left(c_n^2+d_n^2\right)p_n^s(z) + \left(c_{n-1}\,d_n\right)p_{n-1}^s(z) + \left(c_n\,d_{n+1}\right)p_{n+1}^s(z)\right]. \tag{B6a}$$

On the other hand, if we multiply both sides of Eq. (B5b) by $E+M$ for $E < -M$ and use Eq. (B5a), we obtain

$$\left(E^2-M^2\right)p_n^s(z) = \lambda^2\left[\left(c_n^2+d_{n+1}^2\right)p_n^s(z) + \left(c_n\,d_n\right)p_{n-1}^s(z) + \left(c_{n+1}\,d_{n+1}\right)p_{n+1}^s(z)\right]. \tag{B6b}$$

One of the two (B6) relations is associated with spin up and the other with spin down. Both are identical to the three-term recursion relation (4) where we adopt the following parameter assignments

$$z = (E^2-M^2)/\lambda^2\,, \qquad \alpha_n^s = c_n^2 + d_n^2 := \alpha_n^\downarrow\,, \qquad \beta_n^s = c_n\,d_{n+1} := \beta_n^\downarrow\,. \tag{B7a}$$

$$z = (E^2-M^2)/\lambda^2\,, \qquad \alpha_n^s = c_n^2 + d_{n+1}^2 := \alpha_n^\uparrow\,, \qquad \beta_n^s = c_{n+1}\,d_{n+1} := \beta_n^\uparrow\,. \tag{B7b}$$

The recursion coefficients of the spectral polynomials $p_n^{\uparrow\downarrow}(z)$ can be transformed into each other by the parameter map $c_n \mapsto d_{n+1}$ and $d_n \mapsto c_n$. Therefore, $\left\{p_n^s(z)\right\}$ is a sequence of orthogonal spectral polynomials in $z$ that satisfy the symmetric three-term recursion relation (4) with $\left\{\alpha_n,\beta_n\right\} \mapsto \left\{\alpha_n^s,\beta_n^s\right\}$ shown in (B7). The initial values are $p_0^s(z)=1$ and $p_1^s(z) = (z-\alpha_0^s)/\beta_0^s$. Moreover, $\left\{p_n^s(z)\right\}$ satisfy the orthogonality relation (5) with $\rho(z) \mapsto \rho^s(z)$. One can also show that the pair of equations (B4) give

$$-\vec{\nabla}^2\phi_n^+(x) = \lambda^2\left[\alpha_n^\uparrow\phi_n^+(x) + \beta_{n-1}^\uparrow\phi_{n-1}^+(x) + \beta_n^\uparrow\phi_{n+1}^+(x)\right], \tag{B8a}$$

$$-\vec{\nabla}^2\phi_n^-(x) = \lambda^2\left[\alpha_n^\downarrow\phi_n^-(x) + \beta_{n-1}^\downarrow\phi_{n-1}^-(x) + \beta_n^\downarrow\phi_{n+1}^-(x)\right]. \tag{B8b}$$

The conjugate quantum field $\overline{\Psi}^{\uparrow\downarrow}(t,\vec{r})$ is obtained from (B2) by the maps: $\Psi_{\pm}^{\uparrow\downarrow} \mapsto \left(\Psi_{\pm}^{\uparrow\downarrow}\right)^{\dagger}$ and $\phi_n^{\pm} \mapsto \overline{\phi}_n^{\pm}$ where

$$\left\langle \phi_n^r(\vec{r}) \middle| \overline{\phi}_m^{r'}(\vec{r}) \right\rangle = \left\langle \overline{\phi}_n^{r}(\vec{r}) \middle| \phi_m^{r'}(\vec{r}) \right\rangle = \delta_{r,r'}\,\delta_{n,m}\,, \tag{B9a}$$

$$\sum_{n=0}^{\infty} \phi_n^r(\vec{r})\,\overline{\phi}_n^{r'}(\vec{r}\,') = \sum_{n=0}^{\infty} \overline{\phi}_n^{r}(\vec{r})\,\phi_n^{r'}(\vec{r}\,') = \delta_{r,r'}\,\delta^3(\vec{r}-\vec{r}\,')\,, \tag{B9b}$$

where $r$ and $r'$ stand for the four spinor components $1,2,3,4$ with $\phi_n^{+} = \begin{pmatrix} \phi_n^1 \\ \phi_n^2 \end{pmatrix}$ and $\phi_n^{-} = \begin{pmatrix} \phi_n^3 \\ \phi_n^4 \end{pmatrix}$. In $D+1$ space-time, the length scale (conformal degree) of $\phi_n(\vec{r})$ and $\overline{\phi}_n(\vec{r})$ is $-D/2$. If we adopt the conventional notation that $\overline{\phi}_n = \chi_n^{\dagger}\gamma^0$ [i.e., $\overline{\phi}_n^{\pm} = \pm(\chi_n^{\pm})^{\dagger}$] then the orthogonality (B9a) gives $\sum_{r=1}^{4}\left\langle \phi_n^r(\vec{r}) \middle| \chi_m^r(\vec{r})^* \right\rangle = \sum_{r=1}^{4}\left\langle \chi_n^r(\vec{r})^* \middle| \phi_m^r(\vec{r}) \right\rangle = 0$ and we can rewrite (B9) as

$$\left\langle \phi_n(\vec{r}) \middle| \chi_m^{\dagger}(\vec{r}) \right\rangle = \left\langle \phi_m(\vec{r}) \middle| \chi_n^{\dagger}(\vec{r}) \right\rangle = \gamma^0\,\delta_{n,m}\,. \tag{B10a}$$

$$\sum_{n=0}^{\infty} \phi_n(\vec{r})\,\chi_n^{\dagger}(\vec{r}\,') = \sum_{n=0}^{\infty} \phi_n(\vec{r}\,')\,\chi_n^{\dagger}(\vec{r}) = \gamma^0\delta^3(\vec{r}-\vec{r}\,')\,, \tag{B10b}$$

Moreover, we can write $\overline{\Psi}^{\uparrow\downarrow}(t,\vec{r})$ as follows

$$\overline{\Psi}_{\pm}^{s}(t,\vec{r}) = \int dE\sqrt{2E}\,e^{\mathrm{i}Et}\sum_{n=0}^{\infty} b_n^{s\dagger}(E)\,\overline{\phi}_n^{\pm}(\vec{r})\,, \tag{B11}$$

Using the anti-commutators (B3), we can write

$$\left\{\Psi_r^s(t,\vec{r}),\overline{\Psi}_{r'}^{s'}(t',\vec{r}\,')\right\} = \delta_{s,s'}\sum_{n,m=0}^{\infty} \phi_n^r(\vec{r})\,\overline{\phi}_m^{r'}(\vec{r}\,')\int e^{-\mathrm{i}E(t-t')}\rho^s(z)\,p_n^s(z)\,p_m^s(z)\,dz\,, \tag{B12}$$

where we took $f^s(E) \equiv \lambda^{-2}\rho^s(z)$. As in conventional QFT, this defines the singular distribution $\Delta_{r,r'}(t-t',\vec{r}-\vec{r}\,')$ by

$$\left\{\Psi_r^s(t,\vec{r}),\overline{\Psi}_{r'}^{s'}(t',\vec{r}\,')\right\} = \delta_{s,s'}\,\Delta_{r,r'}(t-t',\vec{r}-\vec{r}\,')\,. \tag{B13}$$

Using the orthogonality (5) of the spectral polynomials $\left\{p_n^s(z)\right\}$ and the completeness (B9b) of the pair $\left\{\phi_n^r(\vec{r}),\overline{\phi}_m^{r'}(\vec{r}\,')\right\}$, Eq. (B12) with $t=t'$ becomes

$$\left\{\Psi_r^s(t,\vec{r}),\overline{\Psi}_{r'}^{s'}(t,\vec{r}\,')\right\} = \delta_{s,s'}\,\delta_{r,r'}\,\delta^3(\vec{r}-\vec{r}\,')\,, \tag{B14}$$

Equations (B13) and (B14) give $\Delta_{r,r'}(0,\vec{r}-\vec{r}\,') = \delta_{r,r'}\,\delta^3(\vec{r}-\vec{r}\,')$. Moreover, it is straightforward to write

$$\left\{\Psi_r^s(t,\vec{r}),\Psi_{r'}^{s'}(t,\vec{r}\,')\right\} = \left\{\overline{\Psi}_r^s(t,\vec{r}),\overline{\Psi}_{r'}^{s'}(t,\vec{r}\,')\right\} = 0\,. \tag{B15}$$

Therefore, the canonical conjugate to the spinor quantum field $\Psi_r^s(t,\vec{r})$ becomes $\Pi_r^s(t,\vec{r}) = \mathrm{i}\bar{\Psi}_r^s(t,\vec{r})$.

We can represent the positive-energy spinor particle by the quantum field

$$\mathcal{X}_{\pm}(t,\vec{r}) = \frac{1}{\sqrt{2}}\left[\mathbf{\Psi}_{\pm}^{\uparrow}(t,\vec{r}) + \mathbf{\Psi}_{\pm}^{\downarrow}(t,\vec{r})\right], \tag{B16a}$$

where $\mathbf{\Psi}_{\pm}^{s}(t,\vec{r}) = \Psi_{\pm}^{s}(t,\vec{r}) \pm \hat{\Psi}_{\pm}^{s}(t,\vec{r})^{\dagger}$ and $\hat{\Psi}_{r}^{s}(t,\vec{r})$ is identical to $\Psi_{r}^{s}(t,\vec{r})$ except that $b_n^s(E) \mapsto \hat{b}_n^s(E)$, which are independent annihilation fermion operators associated with the second independent solution of the Dirac equation. The $\pm$ sign to the left of $\hat{\Psi}_{\pm}^{s}(t,\vec{r})^{\dagger}$ comes from multiplication (on the right) of the 4-component spinor $(\hat{\Psi}^s)^{\dagger}$ by the gamma matrix $\gamma^0$. On the other hand, the corresponding anti-particle is represented by the negative-energy quantum field

$$\bar{\mathcal{X}}_{\pm}(t,\vec{r}) = \frac{1}{\sqrt{2}}\left[\bar{\mathbf{\Psi}}_{\pm}^{\uparrow}(t,\vec{r}) + \bar{\mathbf{\Psi}}_{\pm}^{\downarrow}(t,\vec{r})\right]. \tag{B16b}$$

The components of the spectral propagator of the spinor are:

$$\Delta_{n,m}^{\uparrow}(z) = \rho^{\uparrow}(z) p_n^{\uparrow}(z) p_m^{\uparrow}(z), \tag{B17a}$$

$$\Delta_{n,m}^{\downarrow}(z) = \rho^{\downarrow}(z) p_n^{\downarrow}(z) p_m^{\downarrow}(z). \tag{B17b}$$

Due to the anti-commutation relation $\left\{b_0^s(E), b_0^{s'\dagger}(E')\right\} = \delta_{s,s'}\,\lambda^{-2}\rho^s(z)\delta(E-E')$, the mixed propagator components vanish, $\Delta_{n,m}^{\uparrow\downarrow}(z) = 0$. The Feynman propagator is a combined sum of these two in (B17). Had we included both spin components in the interaction model of Section 3, each of the Feynman diagrams shown therein would have consisted of several copies. For example, the self-energy diagram of Fig. 1 would have been the sum of four diagrams obtained from Fig. 1 by the maps:

$$\Delta_{i,k}(z') \mapsto \Delta_{i,k}^{\uparrow}(z'),\ \Delta_{l,j}(z'') \mapsto \Delta_{l,j}^{\uparrow}(z''),\ \eta_n^{i,j} \mapsto \eta_n^{i\uparrow,j\uparrow},\ \eta_m^{k,l} \mapsto \eta_m^{k\uparrow,l\uparrow} \tag{B18a}$$

$$\Delta_{i,k}(z') \mapsto \Delta_{i,k}^{\downarrow}(z'),\ \Delta_{l,j}(z'') \mapsto \Delta_{l,j}^{\downarrow}(z''),\ \eta_n^{i,j} \mapsto \eta_n^{i\downarrow,j\downarrow},\ \eta_m^{k,l} \mapsto \eta_m^{k\downarrow,l\downarrow} \tag{B18b}$$

$$\Delta_{i,k}(z') \mapsto \Delta_{i,k}^{\uparrow}(z'),\ \Delta_{l,j}(z'') \mapsto \Delta_{l,j}^{\downarrow}(z''),\ \eta_n^{i,j} \mapsto \eta_n^{i\uparrow,j\downarrow},\ \eta_m^{k,l} \mapsto \eta_m^{k\uparrow,l\downarrow} \tag{B18c}$$

$$\Delta_{i,k}(z') \mapsto \Delta_{i,k}^{\downarrow}(z'),\ \Delta_{l,j}(z'') \mapsto \Delta_{l,j}^{\uparrow}(z''),\ \eta_n^{i,j} \mapsto \eta_n^{i\downarrow,j\uparrow},\ \eta_m^{k,l} \mapsto \eta_m^{k\downarrow,l\uparrow} \tag{B18d}$$

where $\eta_n^{i\downarrow,j\uparrow} = \left(\kappa/\sqrt{\alpha_n\beta_n}\right)\left[\left(\hat{\alpha}_i^{\downarrow}\hat{\beta}_j^{\uparrow}\right)^{-1} - \left(\hat{\alpha}_j^{\uparrow}\hat{\beta}_i^{\downarrow}\right)^{-1}\right] = -\eta_n^{j\uparrow,i\downarrow}$ and so on. The two diagrams associated with (B18c) and (B18d) are obtained from each other by the exchange of spins making them topologically equivalent and of the same numerical value.

Finally, we like to point out the possibility of an alternative representation of spinors in spectral QFT where matrix-valued orthogonal polynomials are used instead of the scalar (classical) orthogonal polynomials $p_n^{\uparrow\downarrow}(z)$. In that case, the associated spectral propagator $\Delta_{n,m}^{\uparrow\downarrow}(z)$ becomes a matrix rather than a scalar. For an introduction and extensive literature on these polynomials, one may consult [15] and references cited therein. More recent work on the matrix-valued version of the classical Hermite and Gegenbauer polynomials that we used in Appendix A, can be found in [16,17].

## Appendix C: Massless vector field in spectral QFT

In this Appendix, we give a brief description of the massless vector field in spectral QFT. The particle associated with this quantum field (e.g., the photon) has two degrees of freedom corresponding to the two states of polarizations that transverse its propagation, which we designate by the superscripts $\rightleftarrows$. It is constructed using the solution space of the massless Klein-Gordon equation: $\left(\partial_t^2 - \vec{\nabla}^2\right) A_\mu^{\rightleftarrows}(t,\vec{r}) = 0$. The space is reduced from four degrees of freedom carried by $A_\mu^s$ to two by imposing the following two constraints:

(1) Removing the gauge field modes of the form $A_\mu^s = \partial_\mu \Phi^s$ where $\Phi^s$ is a dimensionless space-time scalar function, and

(2) Imposing the Lorenz invariant Landau gauge fixing condition $\sum_{\mu=0}^{\mu=3} \partial^\mu A_\mu^s = 0$,

where *s* stands for the polarizations $\rightleftarrows$.

The positive-energy expansion of the massless vector field is written as follows:

$$A_\mu^{\rightleftarrows}(t,\vec{r}) = \int dE \sqrt{2E}\, e^{-iEt} \sum_{n=0}^{\infty} a_n^{\rightleftarrows}(E)\, \phi_n^\mu(\vec{r}), \tag{C1}$$

where $a_n^s(E) = a_0^s(E)\, p_n^s(z)$ making $p_0^s(z) = 1$ and $\vec{\phi}_n(\vec{r}) \neq \vec{\nabla}\varphi_n(\vec{r})$. The creation/annihilation operators satisfy the commutation algebra $\left[a_0^s(E), a_0^{s'\dagger}(E')\right] = \delta_{s,s'}\lambda^{-2}\rho^s(z)\,\delta(E-E')$ where $\rho^s(z)$ is the positive definite weight function associated with the spectral polynomials $p_n^s(z)$. The Landau gauge fixing, $\partial^\mu A_\mu^s = 0$, becomes $E\phi_n^0(\vec{r}) = -i\vec{\nabla}\cdot\vec{\phi}_n(\vec{r})$ and we require that:

$$\text{For } E>0: \qquad -i\vec{\nabla}\cdot\vec{\phi}_n(\vec{r}) = \lambda\left[c_n\phi_n^0(\vec{r}) + d_n\phi_{n-1}^0(\vec{r})\right], \tag{C2a}$$

$$\text{For } E<0: \qquad -i\vec{\nabla}\cdot\vec{\phi}_n(\vec{r}) = \lambda\left[c_n\phi_n^0(\vec{r}) + d_{n+1}\phi_{n+1}^0(\vec{r})\right], \tag{C2b}$$

where $\{c_n, d_n\}$ are constant parameters such that $c_n d_{n+1} > 0$ for all *n*. These differ from the spinor parameters $\{c_n, d_n\}$ in Eq. (B4). Substituting (C2) in the Landau condition $\partial^\mu A_\mu^s = 0$, we obtain:

$$\text{For } E>0: \qquad E\,p_n^s(z) = \lambda\left[c_n p_n^s(z) + d_{n+1}p_{n+1}^s(z)\right]. \tag{C3a}$$

$$\text{For } E<0: \qquad E\,p_n^s(z) = \lambda\left[c_n p_n^s(z) + d_n p_{n-1}^s(z)\right]. \tag{C3b}$$

For $E<0$, we multiply both sides of (C3a) by *E* and use (C3b) giving

$$(E/\lambda)^2\, p_n^s(z) = \alpha_n^s p_n^s(z) + \beta_{n-1}^s p_{n-1}^s(z) + \beta_n^s p_{n+1}^s(z), \tag{C4}$$

making $z=E^2/\lambda^2$, $\alpha_n^s=c_n^2+d_{n+1}^2$, and $\beta_n^s=c_{n+1}\,d_{n+1}$. On the other hand, for $E>0$, we multiply both sides of (C3b) by *E* and use (C3a) giving the same three-term recursion relation (C4) but with $\alpha_n^s=c_n^2+d_n^2$ and $\beta_n^s=c_n\,d_{n+1}$. We associate the former recursion coefficients $\left\{\alpha_n^s,\beta_n^s\right\}$ with $p_n^{\rightarrow}(z)$ and the latter with $p_n^{\leftarrow}(z)$. Moreover, the free wave equation $\left(\partial_t^2-\vec{\nabla}^2\right)A_\mu^s(t,\vec{r})\;=0$ becomes the three-term recursion relation (C4) dictating that

$$-\vec{\nabla}^2\phi_n^\mu(\vec{r})=\lambda^2\left[\alpha_n^s\phi_n^\mu(\vec{r})+\beta_{n-1}^s\phi_{n-1}^\mu(\vec{r})+\beta_n^s\phi_{n+1}^\mu(\vec{r})\right]. \tag{C5}$$

In addition to the recursion relation (C4), the spectral polynomials $\left\{p_n^s(z)\right\}$ satisfy the following orthogonality relation

$$\int\rho^s(z)p_n^s(z)p_m^s(z)\,dz=\delta_{n,m}, \tag{C6}$$

The conjugate quantum field $\overline{A}_\mu^{\rightleftarrows}(t,\vec{r})$ is obtained from (C1) by complex conjugation and the replacement $\phi_n^\mu(\vec{r})\mapsto\overline{\phi}_n^\mu(\vec{r})$ where

$$\left\langle\phi_n^\mu(\vec{r})\middle|\overline{\phi}_m^\nu(\vec{r})\right\rangle=\left\langle\overline{\phi}_n^\mu(\vec{r})\middle|\phi_m^\nu(\vec{r})\right\rangle=\delta_{\mu\nu}\,\delta_{n,m}, \tag{C7a}$$

$$\sum_{n=0}^{\infty}\phi_n^\mu(\vec{r})\overline{\phi}_n^\nu(\vec{r}')=\sum_{n=0}^{\infty}\overline{\phi}_n^\mu(\vec{r})\phi_n^\nu(\vec{r}')=\delta_{\mu\nu}\,\delta^3(\vec{r}-\vec{r}'). \tag{C7b}$$

Therefore, we can write $\overline{A}_\mu^{\rightleftarrows}(t,\vec{r})=\int dE\sqrt{2E}\,e^{\mathrm{i}Et}\sum_{n=0}^{\infty}a_n^{\rightleftarrows\dagger}(E)\,\overline{\phi}_n^\mu(\vec{r})$ and with the use of the above properties (including the commutator algebra of the creation/annihilation operators), we obtain the equal time commutation relation of the field operators as follows

$$\left[A_\mu^s(t,\vec{r}),\overline{A}_\nu^{s'}(t,\vec{r}')\right]=\delta_{s,s'}\,\delta_{\mu\nu}\,\delta^3(\vec{r}-\vec{r}'). \tag{C8}$$

However, in general, we write $\left[A_\mu^s(t,\vec{r}),\overline{A}_\nu^{s'}(t',\vec{r}')\right]=\delta_{s,s'}\Delta_{\mu,\nu}(t-t',\vec{r}-\vec{r}')$ where

$$\Delta_{\mu,\nu}(t-t',\vec{r}-\vec{r}')=\sum_{n,m=0}^{\infty}\phi_n^\mu(\vec{r})\overline{\phi}_m^\nu(\vec{r}')\int e^{-\mathrm{i}E(t-t')}\rho^s(z)p_n^s(z)p_m^s(z)\,dz. \tag{C9}$$

Thus, $\Delta_{\mu,\nu}(0,\vec{r}-\vec{r}')=\delta_{\mu\nu}\,\delta^3(\vec{r}-\vec{r}')$. The real (neutral) massless particle (e.g., the photon) is represented by the quantum field $\mathcal{A}_\mu(t,\vec{r})=\frac{1}{\sqrt{2}}\left[A_\mu(t,\vec{r})+\lambda^{-1}\overline{A}_\mu(t,\vec{r})\right]$ where $A_\mu(t,\vec{r}):=$ $A_\mu^{\rightarrow}(t,\vec{r})+A_\mu^{\leftarrow}(t,\vec{r})$ and $\overline{\phi}_n^\mu(\vec{r})=\lambda\phi_n^\mu(\vec{r})^*$. However, the complex (charged) massless particle is represented by the positive-energy quantum field $\mathcal{A}_\mu(t,\vec{r}):=\frac{1}{\sqrt{2}}\left[\mathbf{A}_\mu^{\rightarrow}(t,\vec{r})+\mathbf{A}_\mu^{\leftarrow}(t,\vec{r})\right]$ where $\mathbf{A}_\mu^s(t,\vec{r}):=A_\mu^s(t,\vec{r})+\hat{A}_\mu^s(t,\vec{r})^\dagger$ with $\hat{A}_\mu^s(t,\vec{r})$ being identical to $A_\mu^s(t,\vec{r})$ except that $a_n^s(E)\mapsto$ $\hat{a}_n^s(E)$, which is an independent annihilation operator associated with the second independent solution of the massless Klein-Gordon equation. The anti-particle is represented by the negative-energy quantum field $\overline{\mathcal{A}}_\mu(t,\vec{r}):=\frac{1}{\sqrt{2}}\left[\overline{\mathbf{A}}_\mu^{\rightarrow}(t,\vec{r})+\overline{\mathbf{A}}_\mu^{\leftarrow}(t,\vec{r})\right]$. The components of the spectral propagator for a massless vector field are:

$$\square^{\rightarrow}_{n,m}(z) = \rho^{\rightarrow}(z)\, p^{\rightarrow}_n(z)\, p^{\rightarrow}_m(z)\,, \tag{C10a}$$

$$\square^{\leftarrow}_{n,m}(z) = \rho^{\leftarrow}(z)\, p^{\leftarrow}_n(z)\, p^{\leftarrow}_m(z)\,. \tag{C10b}$$

Due to the commutation relation $\left[a_0^s(E), a_0^{s'\dagger}(E')\right] = \delta_{s,s'}\,\lambda^{-2}\rho^s(z)\,\delta(E-E')$, the mixed components of the propagator vanish, $\square^{\rightleftarrows}_{n,m}(z) = 0$. The Feynman propagator is a combined sum of the two in (C10).

A practical initial study in spectral QFT would be to reproduce some of the well-known results in QED where the photon is represented by the quantum field $\mathcal{A}_\mu(t,\vec{r})$ and the associated spectral polynomial could be taken as the Hermite polynomial $\frac{1}{\sqrt{z}} H_{2n+1}(\sqrt{z})$. The electron is represented by the spinor quantum field $\mathcal{X}(t,\vec{r})$ with the associated spectral polynomial taken, for example, as the Laguerre polynomial $L_n^\nu(z)$ or the Gegenbauer (ultra-spherical) polynomial $C_n^\nu(\cos\vartheta)$ where $\cos\vartheta = (z-1)/(z+1)$. The interaction Lagrangian is

$$\mathscr{L}_I(t,\vec{r}) = \eta \bullet \sum_{\mu=0}^{3} \mathcal{A}_\mu(t,\vec{r})[\bar{\mathcal{X}}(t,\vec{r})\gamma^\mu \mathcal{X}(t,\vec{r})]\,, \tag{C11}$$

which is cubic in the field operators similar to the model studied in Section 3.

With $A^{\rightleftarrows}_\mu(t,\vec{r})$ representing the massless gauge vector field that mediates interaction, we can formulate the non-abelian version of spectral QFT by constructing the following quantum field matrices

$$\mathbb{A}^{\rightleftarrows}_\mu(t,\vec{r}) := \sum_{k=1}^{K^2-1} [A^{\rightleftarrows}_\mu(t,\vec{r})]^k\, T_k\,, \tag{C12}$$

where $\{T_k\}$ are the $K^2-1$ Hermitian $K\times K$ matrix generators of SU($K$) with a Lie algebra and normalization

$$\left[T_i, T_j\right] = \mathrm{i}\sum_{k=1}^{K^2-1} C^k_{i,j} T_k\,, \qquad \mathrm{Tr}\left(T_i\, T_j\right) = \tfrac{1}{2}\delta_{i,j}\,. \tag{C13}$$

An example of an interaction Lagrangian in a non-abelian spectral QFT involving the $K$-component spinors and the gauge vector field could read:

$$\mathscr{L}_I(t,\vec{r}) = \eta \bullet \sum_{k=1}^{K^2-1}\sum_{\mu=0}^{3} [\mathcal{A}_\mu(t,\vec{r})]^k [\bar{\mathcal{X}}(t,\vec{r})\left(\gamma^\mu \otimes T_k\right)\mathcal{X}(t,\vec{r})]\,. \tag{C14}$$

## Appendix D: Massive vector field in spectral QFT

In this Appendix, we give a brief description of the massive vector field in spectral QFT. The particle associated with this quantum field (e.g., vector meson) has three degrees of freedom, which we designate as $\vec{V}_\mu(t,\vec{r}) = \left\{V^0_\mu, V^\pm_\mu\right\}$. It is constructed using the solution space of the free wave equation: $\left(\vec{\nabla}^2 - \partial_t^2\right)\vec{V}_\mu(t,\vec{r}) = M^2\vec{V}_\mu(t,\vec{r})$. The space is reduced from four degrees of freedom carried by each of the components $\{\vec{V}_\mu\}$ to three by imposing the Lorenz invariant divergence condition $\sum_{\mu=0}^{\mu=3}\partial^\mu \vec{V}_\mu = 0$.

The positive-energy quantum field associated with the massive vector boson is written as follows

$$\vec{V}_\mu(t,\vec{r}) = \int dE\sqrt{2E}\,e^{-\mathrm{i}Et}\sum_{n=0}^{\infty}\vec{a}_n(E)\,\phi_n^\mu(\vec{r}), \tag{D1}$$

where $\vec{a}_n(E) = \vec{a}_0(E)\vec{p}_n(E)$ making $\vec{p}_0(E) = 1$. The creation and annihilation operators satisfy the following commutation relations

$$\left[a_0^s(E), a_0^{s'\dagger}(E')\right] = \delta_{s,s'}\lambda^{-2}\rho^s(z)\,\delta(E-E'), \tag{D2}$$

where $s$ and $s'$ stand for the three-vector indices $\{0,+,-\}$. The free wave equation $\left(\partial_t^2 - \vec{\nabla}^2 + M^2\right)V_\mu^s(t,\vec{r}) = 0$ becomes a three-term recursion relation for the spectral polynomials $\{p_n^s(z)\}$ dictating that

$$-\vec{\nabla}^2\phi_n^\mu(\vec{r}) = \lambda^2\left[\alpha_n^s\phi_n^\mu(\vec{r}) + \beta_{n-1}^s\phi_{n-1}^\mu(\vec{r}) + \beta_n^s\phi_{n+1}^\mu(\vec{r})\right]. \tag{D3}$$

The divergence condition, $\partial^\mu\vec{V}_\mu = 0$, becomes $E\phi_n^0(\vec{r}) = -\mathrm{i}\vec{\nabla}\cdot\vec{\phi}_n(\vec{r})$ and we require that:

$$\text{For } E > +M: \qquad -\mathrm{i}\vec{\nabla}\cdot\vec{\phi}_n(\vec{r}) = \lambda\left[\left(c_n - \lambda^{-1}M\right)\phi_n^0(\vec{r}) + d_n\phi_{n-1}^0(\vec{r})\right], \tag{D4a}$$

$$\text{For } E < -M: \qquad -\mathrm{i}\vec{\nabla}\cdot\vec{\phi}_n(\vec{r}) = \lambda\left[\left(c_n + \lambda^{-1}M\right)\phi_n^0(\vec{r}) + d_{n+1}\phi_{n+1}^0(\vec{r})\right], \tag{D4b}$$

where $\{c_n, d_n\}$ are dimensionless constant parameters. Substituting (D4) in the divergence condition, we obtain:

$$\text{For } E > +M: \qquad \left(E+M\right)p_n^s(z) = \lambda\left[c_n p_n^s(z) + d_{n+1}p_{n+1}^s(z)\right]. \tag{D5a}$$

$$\text{For } E < -M: \qquad \left(E-M\right)p_n^s(z) = \lambda\left[c_n p_n^s(z) + d_n p_{n-1}^s(z)\right]. \tag{D5b}$$

Therefore, we end up with two possible spectral polynomials that satisfy either one of the following two recursion relations:

$$\left(E^2 - M^2\right)q_n^1(z) = \lambda^2\left[\left(c_n^2 + d_n^2\right)q_n^1(z) + \left(c_{n-1}\,d_n\right)q_{n-1}^1(z) + \left(c_n\,d_{n+1}\right)q_{n+1}^1(z)\right]. \tag{D6a}$$

$$\left(E^2 - M^2\right)q_n^2(z) = \lambda^2\left[\left(c_n^2 + d_{n+1}^2\right)q_n^2(z) + \left(c_n\,d_n\right)q_{n-1}^2(z) + \left(c_{n+1}\,d_{n+1}\right)q_{n+1}^2(z)\right]. \tag{D6b}$$

These two recursion relations must be associated with two out of the three spectral polynomials $\{p_n^\pm(z), p_n^0(z)\}$. So, now the questions are: First, which two out of these three polynomials correspond to (D6a) and (D6b)? Second, how to obtain the third recursion relation? The answer to the first question is easier since in the limit of zero mass, we should recover the two components of the massless vector field treated in Appendix C. That is, as $M \to 0$ we obtain $\vec{V}_\mu(t,\vec{r}) \to V_\mu^{\pm}(t,\vec{r}) = A_\mu^{\rightleftarrows}(t,\vec{r})$. Therefore, we must conclude that the spectral polynomials $p_n^{+}(z) = q_n^{2}(z)$ and $p_n^{-}(z) = q_n^{1}(z)$ with $z = (E^2 - M^2)/\lambda^2$ . To answer the second question, we resort to the hierarchy structure of supersymmetric tridiagonal systems associated with the spectral polynomials $q_n^1(z)$ and $q_n^2(z)$. Recently, Yamani and Mouayn carried out an elegant investigation of such hierarchy structure in [18] according to which, we can write (see Table 2 therein)

$$\left(E^2 - M^2\right) p_n^0(z) = \lambda^2 \left[ \left(c_{n+1}^2 + d_n^2\right) p_n^0(z) + \left(c_n\, d_n\right) p_{n-1}^0(z) + \left(c_{n+1}\, d_{n+1}\right) p_{n+1}^0(z) \right]. \qquad \text{(D6c)}$$

It is interesting to note that the recursion coefficient in relations (D6b) and (D6c) are obtained from each other by the exchange $c_n \leftrightarrow d_n$ making $\beta_n^0 = \beta_n^+$ but $\alpha_n^0 \neq \alpha_n^+$.

The conjugate quantum field and associated particles and anti-particles for these massive vector bosons are constructed in a manner analogous to that in the Appendices above. The components of the spectral propagator for the massive vector field are:

$$\Box_{n,m}^{+}(z) = \rho^{+}(z) p_n^{+}(z) p_m^{+}(z)\,, \qquad \text{(D7a)}$$

$$\Box_{n,m}^{-}(z) = \rho^{-}(z) p_n^{-}(z) p_m^{-}(z)\,. \qquad \text{(D7b)}$$

$$\Box_{n,m}^{0}(z) = \rho^{0}(z) p_n^{0}(z) p_m^{0}(z)\,. \qquad \text{(D7c)}$$

The mixed components of the propagator vanish (i.e., $\Box_{n,m}^{+-}(z) = 0$, $\Box_{n,m}^{+0}(z) = 0$, $\Box_{n,m}^{-0}(z) = 0$). The Feynman propagator is a combined sum of the three spectral propagators in (D7).

## Appendix E: Finiteness of closed-loop integrals in the Feynman diagrams of spectral QFT

In this Appendix, we prove that the value of closed-loop integrals in the Feynman diagrams of spectral QFT (i.e., the "fundamental SAQFT integrals") falls within the interval $[-1,+1]$. As an example, we consider the typical integral

$$I_{i,j,k,l}^{p,q}(a) = \int_0^\infty \rho(x) p_i(x) p_j(x) \omega(x+a) q_k(x+a) q_l(x+a)\, dx\,, \qquad \text{(E1)}$$

where $a$ is a positive real parameter. The maximum absolute value of this integral corresponds to monochrome propagation where $i = j := n$ and $k = l := m$. In that case, we can write

$$\max\left| I_{i,j,k,l}^{p,q}(a) \right| = \zeta_{n,m}^{p,q}(a) = \int_0^\infty \rho(x) p_n^2(x) \omega(x+a) q_m^2(x+a)\, dx\,, \qquad \text{(E2)}$$

The orthogonality of the spectral polynomials (5) gives $\int_0^\infty \rho(x) p_n^2(x)\, dx = 1$ for all $n$. Now, for a function $f(x)$ integrable on the interval $0 \le x < \infty$, we can always write $\int_0^\infty |f(x)|\, dx >$ $\int_a^\infty |f(x)|\, dx$. Therefore, $0 < \int_a^\infty \rho(x) p_n^2(x)\, dx < 1$, which could be written as: $\int_a^\infty \rho(x) p_n^2(x)\, dx =$

$\frac{1}{1+b_n^2(a)}<1$, where $b_n(a)$ is a non-zero real parameter that depends on *n* and *a* with $b_n(0)=1$. Making the replacement $x \mapsto x+a$ in this integral, we get

$$0<\int_0^\infty \rho(x+a)p_n^2(x+a)\,dx=\frac{1}{1+b_n^2(a)}<1\,, \tag{E3}$$

for all *n* and *a*. Now, given two functions $f(x)$ and $g(x)$ that are integrable on the interval $0\le x<\infty$ with respect to the measure $dx$, the Hölder inequality states

$$\int_0^\infty \left|f(x)g(x)\right|dx \le \left(\int_0^\infty \left|f(x)\right|dx\right)\left(\int_0^\infty \left|g(x)\right|dx\right). \tag{E4}$$

If we take $f(x):=\rho(x)p_n^2(x)\ge 0$ and $g(x):=\omega(x+a)q_m^2(x+a)\ge 0$, then this inequality gives

$$\begin{gathered}\zeta_{n,m}^{p,q}(a)=\int_0^\infty \rho(x)p_n^2(x)\omega(x+a)q_m^2(x+a)\,dx\le \\ \left(\int_0^\infty \rho(x)p_n^2(x)\,dx\right)\left(\int_0^\infty \omega(x+a)q_m^2(x+a)\,dx\right)=\left(1\right)\left(\tfrac{1}{1+b_m^2(a)}\right)<1\end{gathered} \tag{E5}$$

proving that the value of the fundamental SAQFT integrals (E1) falls within the range $[-1,+1]$.

The finiteness proof of other fundamental SAQFT integrals, including higher order loops with multiple integration, follows the same procedure that starts with the polynomial orthogonality integral, which is equal to 1. Subsequently, the shifted orthogonality integral is positive but less than 1. Then, we use of the Hölder inequality on the product of those integrals.

## Declarations

**Ethical statement**: The author followed all universally approved ethics rules in conducting this research work.

**Data availability**: No data or material were used in producing the results in this theoretical work.

**Conflict of interest**: The author declares that he has no competing interests relating to this work.

**Funding**: This work received no funding.